\documentclass[nofootinbib,11pt,superscriptaddress,tightenlines]{revtex4-1}
\usepackage{amsmath,amssymb,pgf,pgfarrows,pgfnodes,float,appendix,hyperref}
\usepackage{graphicx}
\usepackage[margin=1in]{geometry}
\usepackage{tikz}
\usepackage[normalem]{ulem}
\usepackage{mathrsfs}
\usepackage{slashed}
\usepackage[compat=1.0.0]{tikz-feynman}
\usepackage{wrapfig}
\usepackage{pgfplots}

\begin{document}
%https://www.overleaf.com/project/6422fc674b0497ae2fad3002
\renewcommand{\figurename}{Figure}
\renewcommand\appendixpagename{\centering \large APPENDIX. CAUSAL DIAMOND OVERLAPS}

\newcommand{\ls}{\ensuremath{l_s}} % String length
\newcommand{\ms}{\ensuremath{m_s}} % String scale
\newcommand{\lP}{\ensuremath{}} % Planck length
\newcommand{\mP}{\ensuremath{m_P}} % Planck mass

\def\eps{\ensuremath{\epsilon}}
\def\del{\nabla}
\def\grad{\nabla}
\def\curl{\nabla\times}
\def\div{\nabla\cdot}
\def\p{\partial}
\def\pbar{\bar{\partial}}
\def\zbar{\bar{z}}
\def\para{{\scriptscriptstyle ||}}
\def\jnodot{{\mbox{\emph{\j}}}}
\def\inodot{{\mbox{\emph{\i}}}}
\newcommand{\tr}{\mathop{\rm Tr}}
\newcommand{\str}{\mathop{\rm STr}}
\def\expec#1{\langle #1 \rangle}
\def\ket#1{| #1 \rangle}
\def\bra#1{\langle  #1 |}
\def\cket#1{|\left. #1 \right)}
\def\cbra#1{\left( #1 \right.|}
\def\slash#1{\ensuremath{\;/\!\!\!\! #1}}
\newcommand{\Rea}{{\mathrm{Re}}}
\newcommand{\Ima}{{\mathrm{Im}}}
\newcommand{\cF}{{\mathcal{F}}}
\newcommand{\cG}{{\mathcal{G}}}
\newcommand{\cH}{{\mathcal{H}}}
\newcommand{\cL}{\mathcal{L}}
\newcommand{\cM}{{\mathcal{M}}}
\newcommand{\cN}{{\mathcal{N}}}
\newcommand{\cO}{{\mathcal{O}}}
\newcommand{\cP}{{\mathcal{P}}}
\newcommand{\cR}{{\mathcal{R}}}
\newcommand{\cS}{{\mathcal{S}}}
\newcommand{\cT}{{\mathcal{T}}}
\newcommand{\cU}{{\mathcal{U}}}
\newcommand{\cV}{{\mathcal{V}}}
\newcommand{\cZ}{{\mathcal{Z}}}
\newcommand{\bR}{{\mathbf{R}}}
\newcommand{\bS}{{\mathbf{S}}}
\newcommand{\bT}{{\mathbf{T}}}
\newcommand{\bZ}{{\mathbf{Z}}}

\newcommand{\bea}{\begin{eqnarray}}
\newcommand{\eea}{\end{eqnarray}}

%Paper-specific macros
\newcommand{\nn}{\nonumber}
\newcommand{\tret}{{t_{\mbox{\scriptsize ret}}}}
\newcommand{\xperp}{{\vec{x}_{\perp}}}
\newcommand{\tp}{{t^{\prime}}}
\newcommand{\xp}{{x^{\prime}}}
\newcommand{\zp}{{z^{\prime}}}
\newcommand{\Xp}{{X^{\prime}}}
\newcommand{\xperpp}{{\vec{x}_{\perp}^{\,\prime}}}
\newcommand{\tpp}{{t^{\prime\prime}}}
\newcommand{\xpp}{{x^{\prime\prime}}}
\newcommand{\zpp}{{z^{\prime\prime}}}
\newcommand{\Xpp}{{X^{\prime\prime}}}
\newcommand{\xperppp}{{\vec{x}_{\perp}^{\,\prime\prime}}}
\newcommand{\Xperppp}{{\vec{X}_{\perp}^{\,\prime\prime}}}
\newcommand{\zt}{{\tilde{z}}}
\newcommand{\tl}{{\tilde{l}}}
\newcommand{\tk}{{\tilde{k}}}
\newcommand{\ts}{{\tilde{s}}}
\newcommand{\tj}{{\tilde{j}}}
\newcommand{\tv}{{\tilde{v}}}
\newcommand{\ta}{{\tilde{a}}}
\newcommand{\tx}{{\tilde{x}}}
\newcommand{\ttau}{{\tilde{\tau}}}
\newcommand{\tPi}{{\tilde{\Pi}}}
\newcommand{\trFsq}{\tr F^2}%{{\mathcal O}_{F^{2}}}
\newcommand{\bphi}{\mbox{$\bar{\Phi}$}}
\newcommand{\vX}{\mbox{$\vec{X}$}}
\newcommand{\vx}{\mbox{$\vec{x}$}}
\newcommand{\Umax}{{U_{\mbox{\scriptsize max}}}}
\newcommand{\Umin}{{U_{\mbox{\scriptsize min}}}}
\newcommand{\Uminsq}{{U^2_{\mbox{\scriptsize min}}}}

\def\O{{\cal O}}
\def\F{{\cal F}}
\def\M{{\cal M}}
\def\N{{\cal N}}
\def\Q{{\cal Q}}
\def\S{{\cal S}}
\def\A{{\cal A}}
\def\R{{\cal R}}
\def\L{{\cal L}}

\def\pd#1{{\bf  \textcolor{blue}{[#1]}}}

\count\footins = 1000
\title{\Large{Quantum theory of three-dimensional de Sitter space}}
\author{Sidan A}
\author{Tom Banks}
\affiliation{Department of Physics and NHETC, Rutgers University, Piscataway, NJ 08854}
\author{Willy Fischler}
\affiliation{Department of Physics and Texas Cosmology Center, University of Texas, Austin, TX 78712}

\begin{abstract}
We sketch the construction of a quantum model of 3 dimensional de Sitter space, based on the Covariant Entropy Principle and the observation that semi-classical physics suggests the possibility of a consistent theory of a finite number of unstable massive particles with purely gravitational interactions.  Our model is holographic, finite, unitary, causal, plausibly exhibits fast scrambling, and qualitatively reproduces features of semi-classical de Sitter physics.  In an appendix we outline some  calculations that might lead to further tests of the model.

\end{abstract}

\maketitle
%\tableofcontents

\section{Introduction}

In previous work on de Sitter (dS) space by two of the present authors \cite{tbwf99} the underlying hypothesis has always been that dS space is an infrared regulator of a scattering theory of particles in Minkowski space. This point of view is problematic in three space-time dimensions because the work of \cite{dhj} shows us that the conventional S-matrix does not exist. Locally flat space-time cannot support a total center of mass energy larger than the Planck mass, and there are other peculiar constraints on would-be Mandelstam invariants.  Every different asymptotic boundary condition for particles changes the asymptotics of space-time.

In this paper we will study $dS_3$, based on the Covariant Entropy Principle \cite{fsb, Bousso:2002ju, tbwf}. The maximal area causal diamond of $dS_3$ is finite, so according to the CEP its Hilbert space must be finite dimensional as well. 
%We will further assume that the variables defining this Hilbert space are organized as solutions of the Dirac equation on the holographic screen of the causal patch, with an eigenvalue cutoff to render the entropy finite. 
The Schwarzschild - de Sitter geometry suggests another important lesson about the % finite diamond
theory: 
localized excitations in the bulk are constrained states of holographic degrees of freedom on the horizon, causing a reduction in the entropy~\cite{Banks:2006rx,Banks:2013fr,Draper:2022xzl,Morvan:2022ybp}.  Generic states of the holographic degrees of freedom, with respect to the density matrix of the diamond, carry energy (as measured along the geodesic in the diamond) that scales like $1/R$ as the diamond radius goes to infinity. Bulk localized energy must satisfy $ER\gtrsim 1$ and $ER$ is proportional to the number of constraints on the boundary q-bits.\footnote{In the $RM_P \rightarrow\infty$ limit, the boundary states might seem to become ``zero momentum massless particles".   This is probably an inaccurate description.  In a generic causal diamond, massless particles with momentum of order the diamond radius would generally have wave functions spread throughout the bulk.  The states that contribute the bulk of the entropy have wave functions localized infinitely close to the boundary and give rise to a UV divergence in the entropy.}
 The fact that localized bulk excitations have low entropy suggests that they are all unstable. 

In the first part of this paper we examine the instability of localized excitations from the bulk semi-classical point of view. Given a fixed static patch, instability is visible already in the classical limit, for any excitation whose center of mass follows a geodesic other than the one that connects the time-like tips of the diamond.\footnote{The theory of classical relativistic particles coupled to $3D$ gravity in de Sitter (dS) space was exactly solved by Deser and Jackiw \cite{dj}, following their work with 't Hooft on the analogous problem in locally asymptotically flat space-time \cite{dhj}. A comprehensive study of the  solutions in flat and AdS spaces can be found in the paper by Matschull \cite{matschull}.  The latter work makes it abundantly clear that the gravitational field has no independent degrees of freedom and merely modifies the structure of the particle phase space.} These states all have lifetimes roughly equal to the proper time it takes the geodesic to exit the diamond.  Recall that this is independent of the mass of the particle. In the proper time along any trajectory connecting the time-like tips of the diamond, the excitation fades into the generic background equilibrium of the horizon. Only a particle on the geodesic connecting the tips of a given static patch is classically stable.

Quantum mechanics modifies this classical behavior in three ways:
\begin{itemize}
\item The uncertainty principle limits the localization of the object and its 
velocity fluctuations.  The probability that it follows the static geodesic for an infinite time is zero.
\item In $d \geq 4$ Gibbons-Hawking radiation will subject the object to momentum kicks, which would cause it to random walk toward the horizon.  Once it gets sufficiently far, the further time to exit the horizon is $a R$, with a mass independent coefficient.  
\item If the localized object is a black hole, it will decay by Hawking radiation and the massless radiation will escape to the horizon.  $d =3$ pure gravity has no particle excitations, so it's not clear that every model must have Gibbons-Hawking radiation. Furthermore, the static solutions of classical Einstein gravity in $d=3$ with nonnegative cosmological constant do not have horizons apart from the cosmological one.
\end{itemize} 

The first two sources of instability are minimal if the mass is large, and if the mass is large enough in $d \geq 4$ the object is surely a black hole, or will become one eventually.  Since black holes can have masses of order $R^{d-3}$ there are many long lived systems, gravitationally bound to the origin, which will have lifetimes $\gg R$.  Furthermore, we will argue below that for a large gravitationally bound object with many semi-classical subsystems, the mathematical description of the system from the point of view of a geodesic in dS space is quite irrelevant.  Different trajectories of the center of mass of the system decohere
and the subsystems can use these to define a reference frame in which to do quantum mechanics.  By its definition, this reference frame is stable until the entire object collapses into a black hole and begins to emit Hawking radiation.

In $d = 3$, the situation is quite different.  The maximal mass of a localized object in $dS_3$ (or three dimensional Minkowski space for that matter) is of order the Planck mass.  Roughly speaking then, the wave function of a particle starting out at rest at global time $t = 0$ in $dS_3$, is spread over a distance of order at least $1$ in Planck units. Then the nature of the instability in 3 dimensions is dominated by the first of the mechanisms above.
%: the mass of any localized object in $dS_3$ is less than the Planck mass, so there is a position uncertainty which cannot be made very small. 

In Sec.\ref{sec:summary} we explain and give a summary of previous results. In Sec.~\ref{sec:quantumparticles} we examine the evolution of particle wave functions in $dS_3$.  By examining the transition amplitudes for single particle motion, we  argue that a localized wave function becomes uniformly spread over the static patch {\it coordinates} in a time of order the dS Hubble scale, $R$.  This means that the probability distribution is concentrated near the horizon in that time.  We  interpret this as an instability of the quantum state corresponding to a particle on the static geodesic to decay to a ``typical" thermal state of the vacuum ensemble\footnote{Typicality is meant in the sense of statistical mechanics and the Eigenvalue Thermalization Hypothesis.  The time for the state to become Haar random is much longer.}.  Particles localized on geodesics other than the static trajectory connecting the two tips of the static patch are, of course, classically unstable and have lifetimes of order $R$ as well.  We exhibit the geometrical mechanism by which the ``deficit angle" of such a moving particle disappears from measurements made by a detector localized on the static geodesic.

%In this paper we will use these results to take steps toward constructing a model of quantum gravity in three dimensional dS space.   %We will argue that it is consistent with and provides further evidence for the contention that localized objects are constrained states of holographic degrees of freedom.   

In Sec.\ref{sec:models}, we discuss a set of finite dimensional quantum mechanical models which have the qualitative properties of the semi-classical dynamics outlined above.  The variables in these models are fermion fields $\psi (\sigma,\theta)$ on the holographic circle $\theta$ times a near-horizon interval $\sigma$.  There are cutoffs on both the angular momentum in the $\theta $ direction and the linear momentum on the interval $\sigma$. The cutoffs are correlated in a way we will describe below. States with a deficit angle are constrained by 
\begin{equation} \psi (\theta_n,\sigma) | s \rangle = 0 , \end{equation} if $n$ is a multiple of an integer $p$.  $\theta_n$ are a discrete set of angles such that the fields $\psi(\theta_n, \sigma )$ are independent canonical fermion fields.    There are many choices of Hamiltonian that make all such states decay in a time of order $R$, and have other desirable gravitational attributes, like fast scrambling \cite{Susskind:2005js,Hayden:2007cs,Sekino:2008he}.   However, a natural mechanism for fast scrambling, the ``fuzziness" of higher dimensional geometry, does not work for the one dimensional boundary of $dS_3$, and we have to insert explicit long range interactions into the boundary Hamiltonian to achieve fast scrambling of quantum information.
The ideas of Carlip \cite{carlip} and Solodukhin \cite{solo} provide a guide to finding the right description of these interactions.

\section{Summary of Previous Results}\label{sec:summary}

The basic principles on which this paper is based are the assertion \cite{ted95,fsb} that the Bekenstein-Hawking area law for black holes holds for an arbitrary causal diamond in any model of quantum gravity
\begin{equation}
    \langle K_{\diamond} \rangle = \frac{A_{\diamond}}{4G_N} , 
\end{equation}
and that states localized on a geodesic in de Sitter space have an entropy deficit
\begin{equation}
    \Delta S = 2\pi R E .
\end{equation}
Here $R$ is the dS radius and $E$ is the energy in static coordinates.  In four and higher dimensions, this formula is only valid to the leading order when $ER$ is small.  The strongest argument for the first assertion is Jacobson's demonstration \cite{ted95} that the hydrodynamic equations of this entropy law are the double null projections of Einstein's equations.  It can also be demonstrated by a Euclidean path integral argument \cite{bdf} or by the generalization to arbitrary diamonds \cite{BZ} of the results of Carlip \cite{carlip} and Solodukhin \cite{solo}.   The latter authors argued that the effective modular Hamiltonian\footnote{The modular Hamiltonian is minus the logarithm of the density matrix.} on the horizon of a black hole was the $L_0$ generator of a (cut-off) $1 + 1$ dimensional conformal field theory, with central charge proportional to the horizon area.  The CFT lives on an interval, the stretched horizon, and its target space is related to the geometry of the holographic screen of the diamond\footnote{The maximal area surface in a null foliation of the boundary of the diamond.}.

The second assertion follows from the first and from the formula for the Schwarzschild dS black hole metric.  It gives a derivation of the Gibbons-Hawking temperature of dS space that depends only on the area law, rather than the details of quantum field theory.  It tells us that static energy is a measure of the number of constrained q-bits in a localized state.   The localized state density matrix is the projection of $e^{-L_0}$ for empty dS space onto a lower dimensional subspace.   This explains why dS space has a fixed temperature.  

In a number of papers, two of the authors (T. B. and W. F.) have suggested that the proper variables for describing the fluctuating geometry of the holographic screen of any causal diamond were the expansion coefficients of a spinor field in eigenspinors of the Dirac equation on the background geometry.  The idea is that the background geometry is a hydrodynamic variable.  Connes \cite{connesncgeom} showed that Riemannian geometry was entirely encoded in the Dirac operator on the Riemannian manifold. The coefficients of the expansion of a solution of the fluctuating geometry's Dirac equation in eigenspinors of the background are a finite set of variables if we impose a UV cutoff on the Dirac eigenvalue.   If they are quantized as (cut-off) $1 + 1$ dimensional fermion fields, following the prescription of Carlip and Solodukhin, then we can correlate the UV cutoff on the transverse Dirac eigenvalue with the central charge of the CFT, and thus with the area of the holographic screen of the diamond. By contrast, the UV cutoff on the $1 + 1$ dimensional Dirac momentum is associated with the very smallest diamonds.  Our interpretation of this is that the Carlip-Solodukhin analysis is semi-classical/hydrodynamic and breaks down for very small systems.  A theoretical lower bound on its use is that the single fermion CFT must have enough states for Cardy's formula to provide a good approximation to the spectrum.  Probably of order 10-20 momentum eigenstates suffices for that.  In the real world, one would have to determine the correct value of the $1 + 1$ dimensional cutoff by experiment.  In the imaginary world of $2 + 1$ dimensional dS space we would have to realize this space-time as a subsystem of some rigorously defined asymptotically flat or AdS system, perhaps along the lines of \cite{evaetal1,evaetal2}, and hope that the larger system contains a sufficiently precise copy of the $dS_3$ subsystem to determine these delicate UV details.  As appears to be the case for linear dilaton gravity in $1 + 1$ dimensions, there might be many such embeddings, and no unique answer to the question.

The authors of \cite{Susskind:2005js, Hayden:2007cs, Sekino:2008he} have argued that the quantum systems on stretched horizons are fast scramblers of quantum information.  T. B. and W. F. have argued that the existing evidence for fast scrambling, for horizons of dimension $2$ or higher,  is accounted for if the systems are invariant under something like the group of volume preserving diffeomorphisms of the horizon, or some fuzzy deformation of it.   $dS_3$ has a horizon of dimension $1$ and length preserving diffeomorphisms preserve order.  Thus any finite version of this group will be a $Z_N$ translation group and the resulting system will not have fast scrambling unless it has long range interactions.  In the text, we've instead written random couplings between different points related by the $Z_N$ symmetry.  We imagine that different instances of these random couplings might be realized from different realizations of the model in AdS/CFT.

\section{Quantum Mechanics of Single Particles in $dS_3$}
\label{sec:quantumparticles}
In quantum field theory, spin zero particles in $dS_3$ are described by the Klein-Gordon equation.  We can view this as single particle Hamiltonian evolution for particles with positive energy in some particular static patch.  Generic local interactions between particles force us to give up the single particle picture, because a transition amplitude in one frame can be viewed as a particle production amplitude in another.   This is not the case for gravitational interactions in $d = 3$.  As shown by Matschull \cite{matschull}, the effect of gravitation in three dimensions is merely to modify the phase space of a fixed number of particles.   While this argument is valid also for negative cosmological constant, no one has found a quantum theory dual to pure AdS gravity coupled to a finite number of particles.  Note also that the static solutions of GR in 3 dimensions with asymptotically AdS boundary conditions have finite area horizons.  The overwhelming evidence from the AdS/CFT correspondence is that the entropy/area law is valid and that those solutions correspond to complicated quantum systems with large Hilbert spaces.   This is not true for $dS_3$ or Minkowski space\footnote{The entire concept of asymptotically flat space is problematic in $3$ dimensions.   Classical ``scattering" solutions exist for point particles \cite{dhj}, but traditional Mandelstam invariants are bounded and there is no obvious Hilbert space on which a traditional S-matrix could be defined.}.   The static solutions have an entropy deficit, just like higher dimensions, but no entropy of their own.  It is certainly impossible to put an infinite number of particles into $dS_3$ if each particle has any finite energy, and, at the classical level, no pair of particles remains in causal contact for more than a finite amount of proper time.  

Our proposal, at the semi-classical level, is to study the quantum mechanics of individual particles. This model does not explain the entropy of empty $dS_3$ or the entropy deficits of localized classical particles, but it does demonstrate the quantum mechanical decay back to the classical equilibrium state represented by empty $dS_3$.  We then propose a quantum model with a finite dimensional Hilbert space that does explain the entropic properties of solutions.  The localized solutions are realized as constrained states of the underlying variables, and the decay rates of those states back to equilibrium agree qualitatively with the semi-classical results.

\subsection{Classical Particles in $dS_3$}

Although the physics of dS space can all be understood in a single static patch, it is convenient for calculations to work in the flat slicing metric
\begin{equation}\label{eq:flatslicing}
ds^2 = - dt^2 + e^{2t/R} d\vec{y}^2 \end{equation} through coordinate transformations from the static patch coordinates $(t_s, r_s)$ to the flat slicing coordinates $(t, r)$
\begin{equation}
    \begin{split}
        r_s &= r e^{t/R} ,\\
        e^{-2t_s/R} &= e^{-2t/R} - r^2/R^2,
    \end{split}
\end{equation}
where $r^2\equiv \vec{y}^2$. The geodesic equations can be solved by writing the conservation law resulting from spatial translation in these coordinates
\begin{equation}   \bigg(1 - e^{2t/R} \frac{d\vec{y}^2}{dt^2}\bigg)^{-1/2} \frac{d\vec{y}}{dt} = \vec{v}_0 . \end{equation}
The solution is
\begin{equation}\label{eq:time-likesolution}
\vec{y} (t) = \vec{y}_0 - \frac{\vec{v}_0} {v_0^2}\bigg(\sqrt{1 - v_0^2} - \sqrt{1 - v_0^2 e^{-2t/R}}\bigg) . \end{equation}

The classical geometry produced by a particle of mass $m$ traveling on this geodesic is obtained from that of $dS_3$ by modding out by an $SO(1,3)$ transformation that leaves the geodesic invariant.  For geodesics with $\vec{v}_0 = 0$, this is just a rotation around the point $\vec{y}_0$, producing a deficit angle $\delta = 8\pi G_N m$ \cite{dj} .   Let's consider a particle at $\vec{y}_0$, viewed from the static patch centered at $\vec{y} = 0$. Take a causal diamond of finite proper time centered on the $\vec{y} = 0$ geodesic with future tip $t_2$ and past tip $t_1$.   The particle enters the causal diamond at flat slicing time 
% \begin{equation} t_i = -\lim_{t_1\rightarrow -\infty}R\ln \bigg(e^{-\frac{t_1}{R}}-\frac{y_0}{R}\bigg) ,\end{equation} and leaves it at
% \begin{equation} t_f = - \lim_{t_2\rightarrow \infty} R\ln \bigg(e^{-\frac{t_2}{R}}+\frac{y_0}{R}\bigg).\end{equation} 
\begin{equation} t_i = -R\ln \bigg(e^{-\frac{t_1}{R}}-\frac{y_0}{R}\bigg) ,\end{equation} and leaves it at
\begin{equation} t_f = - R\ln \bigg(e^{-\frac{t_2}{R}}+\frac{y_0}{R}\bigg).\end{equation} 

\begin{figure}
    \includegraphics[scale = 3]{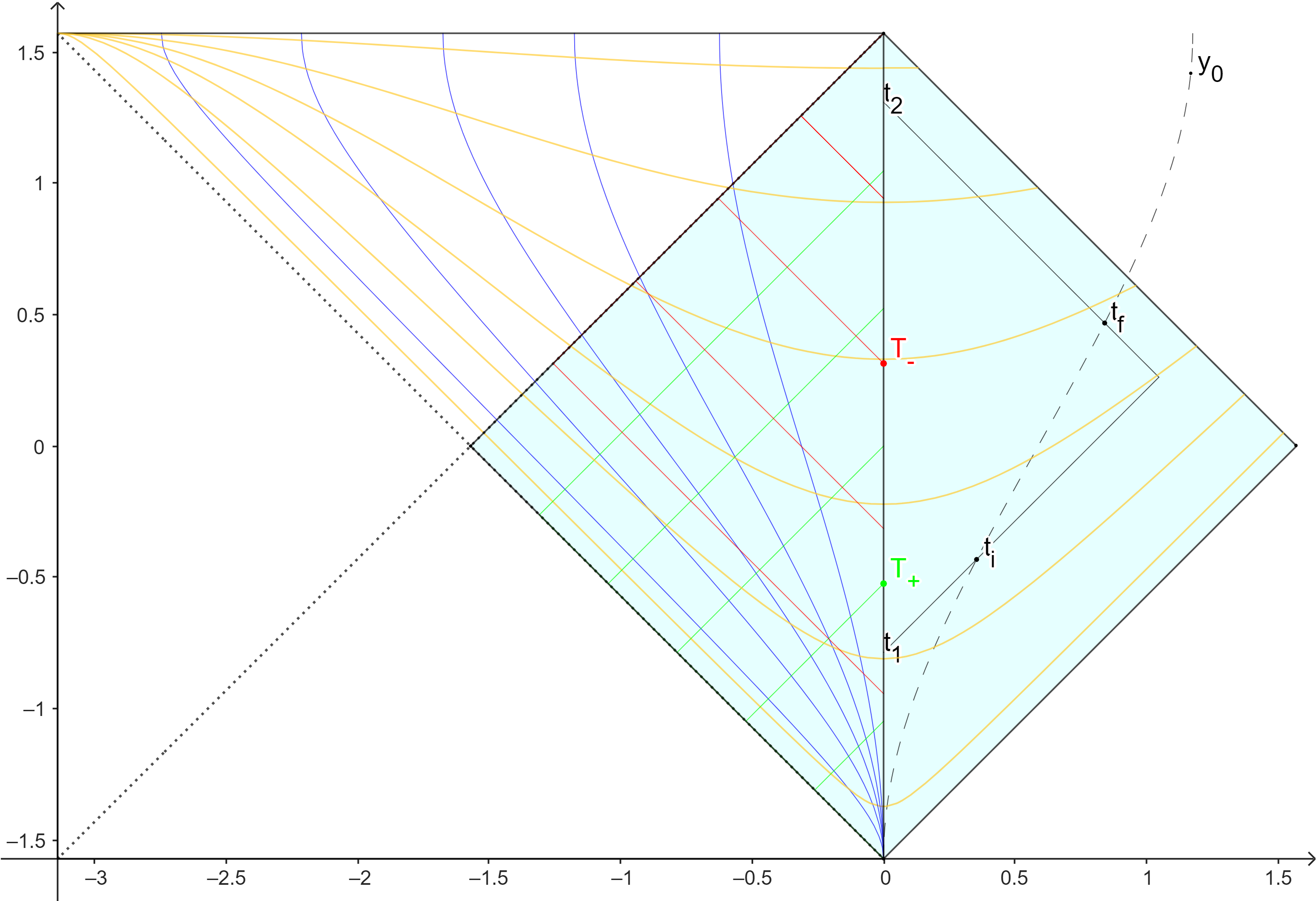}
    \caption{The Penrose diagram of de Sitter space with flat slicing coordinates contours. Yellow curves are constant-$t$ surfaces and blue curves are constant-$y$ surfaces in flat coordinates. The static patch is highlighted in light blue. Green (red) lines are boundaries of future (past) oriented nests of diamonds centered at the origin. The dashed curve is the trajectory of a massive particle at $\Vec{y}_0$ that enters and leaves the causal diamond at time $t_i$ and $t_f$, respectively.}
    \label{fig:nestedcd} 
\end{figure}

Consider past and future oriented nests of diamonds, all centered at the origin (Fig. \ref{fig:nestedcd}).  The entire effect of the particle is encoded in the lengths of circles at fixed flat time $t$.  If a circle winds around the geodesic at $\vec{y}_0$ its circumference is smaller than it would have been in the absence of the particle.  Note that the flat slicing time is equal to the proper time along both geodesics.  If we look at future oriented diamonds with future tip $T_+$ such that
\begin{equation} e^{T_+/R} \int_{-\infty}^{T_+} dt e^{-t/R} < t_i , \end{equation} or past oriented diamonds with past tip $T_-$ such that
\begin{equation} e^{T_- /R}  \int_{T_-}^{\infty} dt e^{-t/R}  > t_f , \end{equation} then there are no circles in those diamonds that circumnavigate the geodesic at $\vec{y}_0$.  In both cases, a detector in the diamond can receive signals from detectors that circumnavigate the particle trajectory, but cannot send signals back.   The interpretation of this for the early, future-pointing nested diamonds is that the particle is part of the initial data on the past boundary of a larger diamond in the nesting.  Since the particle is moving slower than light, the early detectors can receive signals of its imminent arrival\footnote{This phrase uses the idea that there are things outside the static patch.  From the framework of static patch holography, ``imminent arrival" just means that the dynamics in the patch respects causality, in the sense that some of the initial conditions that are necessary to describe the full history of the diamond are imposed on degrees of freedom that are not in the algebra defining a smaller diamond in the nest.}, but cannot perform their own experiments to verify the shortened circles around the particle trajectory.

The question we want to answer is how this classical physics fits into the proposal of \cite{tbwf} that all of the physics of dS space is captured by a finite dimensional system describing a single static patch. A classical relativist is apt to say that the static coordinate system is not good because it doesn't cover the manifold.   A hypothetical quantum system defined on the full flat slices would certainly argue that on flat slices there are circles at all times, which exhibit a length deficit that is not redshifted.  The association of an entropy deficit with a length deficit makes this a challenge for static patch holography.  In the classical picture, there is a subsystem dual to the causal diamond in Fig. \ref{fig:nestedcd}, in which the entropy deficits exist for the entire period that the particle ``remains in the static patch."\footnote{Here period is measured in terms of proper time along the particle's geodesic.}  Unlike the length along space-like circles, an entropy deficit cannot be redshifted away by going to another reference frame.  The quantum principle of relativity says that the density matrix corresponding to the diamond of Fig. \ref{fig:nestedcd} must have the same spectrum computed along the geodesic in the static patch as it does when computed along the geodesic along which the particle travels.  In order for static patch holography to work, the low entropy particle state must decay back to the static patch equilibrium state.

\subsection{Quantum Particles in $dS_3$}\label{sec:IIIB}
Now we consider the evolution of the support of scalar particle wavefunctions in $dS_3$. This provides a bulk semiclassical picture for how the low entropy particle state decays back into the equilibrium state.

In flat polar coordinates the horizon is located at $r = R e^{-t/R}$ where $R$ is the dS radius, and $r$ the coordinate distance from the origin. The general massive scalar solutions are
\begin{align}
\phi_{k \ell}(t,r,\theta) = e^{i \ell \theta} e^{-t/R}  J_\ell(k r) \left( A_{\ell k}\, J_n(k R e^{-t/R}) + B_{\ell k} \,J_{-n}(k R e^{-t/R}) \right)
\end{align}
where $n = \sqrt{1 - R^2 m^2} \approx i m R$.

In the neighborhood of static time $t_s= 0$ and static radius $r_s= 0$,  the flat $t,r$ coordinates are also close to the origin.  Therefore a lump, localized within a flat radius of about $1/m$ of the origin at $t=0$, has most of its support in flat radial wavenumbers $k < m$.  The radial part of the solution above, $J_{\ell}( k r )$, has its first zero at $r \sim 1/k > 1/m$. So the horizon reaches the first Bessel zero for every relevant $k$ at flat times $t < t_{max} = R \log (m R)$. At later flat times, the solution inside the horizon is approximately homogeneous but exponentially damped like $e^{-t/R}$. (The time dependent Bessel functions $J_{\pm imR}(k R e^{-t/R})$ just oscillate with increasing $t$.) 

Static time slices $t_s$ intersect flat slices $t$ iff $t_s > t$, and they do so at static coordinate radius $r_X(t,t_s)$ given by
\begin{align}
r_X = R \sqrt{1-e^{-2 (t_s-t)/R}}.
\end{align}
 For all smaller static radii $r_s<r_X$, the static time slice samples the solution at {\emph{later}} flat times, where the solution is more exponentially suppressed.  So for static times $t_s > t_{max} + ({\rm few})\times R$, the solution in static coordinates is localized near $r_X$, and $r_X$ is exponentially close to the horizon.
 
 Note that there are two free mode coefficients $A_{\ell k}, B_{\ell k}$ in the solution above.  One linear combination of them is determined by the initial condition at time $t_s = t = 0$, and we have analyzed the generic behavior for solutions localized within $1/m$ of the origin at that time.  The other linear combination is determined by the requirement that the solution contains only positive frequencies with respect to static time.  Since the considerations of the previous paragraph were independent of the choice of $A$ and $B$ for a given $\vec{y}$ space initial condition, the positive frequency solutions of the KG equation in static coordinates, corresponding to the quantum behavior of a particle of mass $m$ localized on the static geodesic, will predict overwhelming probability for the particle to be smeared over the horizon in a static time of order $R \, {\rm ln}\ mR$.  Recognizing that $R$ is the natural timescale for interactions on the horizon, this is consistent with the time for a fast scrambling system to equilibrate an entropy deficit of order $mR$.
 
Now let's consider a particle localized on the geodesic at $\vec{y}_0$ with $\vec{v}_0 = 0 $.  The momentum space solutions are exactly the same as the ones above.  We just have to Fourier transform them w.r.t. $\vec{y} - \vec{y}_0$.  Let's consider initial conditions with vanishing angular momentum around $\vec{y}_0$.  The resulting wave function remains circularly symmetric around $\vec{y}_0$ for all time.  We can repeat the calculation above for a light particle initially localized near the static geodesic at $\vec{y}_0$.  The position space wave function is:
\begin{equation} 
    \phi (\vec{y}) = e^{-t/R}\int d^2 k e^{i\vec{k}\cdot(\vec{y} - \vec{y}_0)} 
    [A(k) J_{n} (kR e^{-t/R}) + B(k) J_{-n} (kR e^{-t/R})] . 
\end{equation}    
If $|y_0|$ is large then the value of the initial wave function in the static patch centered at the origin is exponentially small.  As we bring $\vec{y}_0$ closer to the origin, it gets larger and is concentrated in the angular direction of $\vec{y}_0$.   However, the dominant effect we discussed for a particle sitting near the origin remains the same.   The static patch at the origin shrinks exponentially in $y$ coordinate size, but the wave function converges to
 \begin{equation} \phi (\vec{y}) = e^{-t/R}\int d^2 k e^{i\vec{k}\cdot(\vec{y} - \vec{y}_0)} [A(k) J_n (0) + B(k) J_{-n} (0)] . \end{equation} $y_0$ is now outside the static patch centered on the origin. Inside that patch, the solution in flat coordinates is again approximately homogeneous and exponentially damped in time, and so the dominant support in static coordinates is concentrated near the horizon, and the angular dependence disappears. The component of this solution with angular momentum $\ell$ around the origin is proportional to $J_{\ell} (m |y_0|) J_{\ell}(m |y|)$.  Since the horizon moves to exponentially smaller values of $y$ all components with non-zero $\ell$ die off exponentially.  Again we see return to the equilibrium state.  For larger $y_0$ this happens more rapidly because the amplitude of the initial condition is small to begin with.

There is a subtle point in the analysis above. It is well known that the space of solutions of the Klein-Gordon equation is not a Hilbert space, but we have treated the KG solutions as quantum wavefunctions with the standard probability interpretation. This problem is actually resolved in dS space, if we remain within a single static patch. The positive frequency solutions of the equation {\it do} form a Hilbert space in the static patch.  Positive static patch frequency wave functions are complicated linear combinations of the flat coordinate solutions that we have discussed. It's therefore important that our discussion did not depend on particular choices of the functions $A(k)$ and $B(k)$ but only their general support in momentum space.   
 
Finally, we have  to address the question of why this same discussion does not apply in higher-dimensional de Sitter space.   The answer has two parts.   The first is that the Nariai bound on the mass of localized objects in dS scales like $R^{d-3}$ in $d$ dimensions, so the wave functions of localized objects can be much more localized and it takes longer for them to dissipate.   In addition, higher dimensional gravity produces gravitational bound states so that even light objects like electrons can have wave functions that remain localized in a static patch for much longer than the dS time $R$, because they are bound to larger masses.   In our own universe, if the initial conditions are such that a galaxy can form, then the static patch in the rest frame of that galaxy will have localized objects in it until quantum fluctuations {\it of the galaxy center of mass position} predict a very high probability for the galaxy wave function to be spread uniformly over the horizon.  For detectors bound to the galaxy, the lifetime is even longer.   It's not until the galaxy collapses to form a black hole that these detectors ``cease to exist."   The parts of the galaxy positional wave function that follow different time-like trajectories decohere from each other because the galaxy has so many quantum states for each position, and makes transitions between them on a time scale much shorter than the time it takes the center of mass to move.   So the prediction of QM for a detector inside the galaxy is that it will find itself carried along on some time-like trajectory, after which observations of things bound to the galaxy are more or less unaffected by the fact that the galaxy is in dS space.  Particles radiated by the galaxy will disappear into the Gibbons-Hawking haze in a time of order $R$, and it's only when that dissipation leads the galaxy to gravitationally collapse that the detector is affected by the dS background.  Only cosmological observations of objects in the causal past of the galaxy, which are fading into the haze, can detect the global geometry of space-time.    In three space time dimensions there are no large masses and no complex bound systems that can exhibit this kind of interesting long term behavior.   The existence of horizons for the static solutions of GR with non-negative cosmological constant. in dimension greater than three is an indication that this kind of complexity cannot be avoided in any model of quantum gravity.  Although the internal dynamics of black holes does not provide us with many semi-classical observables, there are things like local rates of Hawking decay that are correlated with the center of mass position of the black hole and can be argued to have a decoherent, if evanescent, semi-classical existence.

The three dimensional picture might be modified if we added electromagnetic interactions to the model since preliminary estimates suggest we could make a range of complex atoms and molecules with masses less than the Planck mass and sizes smaller than the dS radius.  However, such a model raises a number of issues. Among these is the inevitability of particle production and how to make that compatible with a bound on the total energy in the system.  In addition, the authors of \cite{Emparan:2022ijy} have suggested that in the presence of massless particles the static solutions of $dS_3$ gravity will develop genuine horizons and large entropy, so that our discussion would have to be modified drastically.  We leave the exploration of models with other light fields to future work.

\section{Towards a Quantum Model of $dS_3$}
\label{sec:models}
The basic strategy of HST models of quantum gravity is to describe the variables on the holographic screen of a diamond as cut off quantum fields on the screen, with a cutoff that preserves the symmetries of the diamond.  The requirement that the diamond has finite entropy means that the basic fields are fermions\footnote{Any finite dimensional Hilbert space is a representation of a one (discrete) parameter family of superalgebras since the Gell-Mann matrices are closed under both commutation and anti-commutation.  Moreover, even if the dimension of the Hilbert space is not a power of $2$, one can realize it as a constrained subspace of the representation space of a finite number of canonical fermions.}.  
The spin-statistics connection and the observation of Connes \cite{connesncgeom}  that Riemannian geometry is encoded in the Dirac operator, suggests that we view them as belonging to the spinor bundle over the screen.  The work of Carlip \cite{carlip} and Solodukhin \cite{solo} suggests that they be viewed as variables in a $1+1$ dimensional CFT, whose $L_0$ generator is the modular Hamiltonian of causal diamonds in $dS_3$ \cite{CHM, JV}.  

Let us elaborate a bit on this last point.  Consider a causal diamond defined by a geodesic with proper time interval $[-T,T]$.  There is a coordinate system $(\tau,x)$ on this diamond, which we will call Diamond Universe (DU) coordinates, where the metric takes the form
\begin{equation}
    \begin{split}
        ds^2 &= C^2 (\tau, x) (-d\tau^2 + dx^2 + \sinh^2 x d\phi^2 ),  \\
        C &= \frac{R\sinh(R_*/R)}{\cosh(\tau) + \cosh (x) \cosh (R_*/R)},
    \end{split}
\end{equation}
where $\tau \in (-\infty, \infty)$ is a time coordinate and $x \in (0, \infty)$ is a radial coordinate. The relation between $T$ and $R_*$ is 
\begin{equation}
    T = \frac{\pi R}{2} \tanh(R_*/R) .
\end{equation}

There is a similar coordinate system for each of the sub-diamonds corresponding to the intervals $[-T, \tau]$.   In each of these systems, the lines of constant $x$ are flow lines of a vector field $V_0 (\tau)$, which leaves the diamond invariant and is a conformal Killing vector (CKV) of $dS_3$.  The generalization of the conjecture of \cite{carlip, solo} made in \cite{BZ} is that in models of quantum gravity, the quantum operator implementing the action of the CKV on the bifurcation surface of each of these diamonds is the $L_0$ generator of a $1+1$ dimensional CFT with central charge proportional to the length of the bifurcation surface.  Connes' results, which motivated the HST formalism, suggest that this CFT be built from
fermion fields $\psi_n (\sigma^0, \sigma)$ labeled by eigensections $\chi_n (y)$ of the $1$ dimensional Dirac operator on the bifurcation surface,
\begin{equation} \gamma e^{-1}(y) i\partial_y \chi_n = \lambda_n \chi_n , \end{equation} where $e(y)$ is the einbein on the boundary and $\gamma$ is one of the Pauli matrices. $y$ is the boundary coordinate and $\sigma$ is a spatial coordinate on an interval where the model is defined.

The restriction to finite entropy implies that we must impose cutoffs on both the transverse Dirac operator and the spectrum of $L_0$.    We will require that the $1 + 1$ dimensional cutoff be independent of the size of the diamond, so that the entropy scaling is determined by the transverse cutoff.  
As we'll see, the cutoffs are correlated, but depend on interactions that we must add to the free $1 + 1$ fermion Lagrangian in order to satisfy some important properties of a sensible model of quantum gravity. These properties have to do with the time evolution operator.   Time evolution in the DU coordinates of necessity involves a time dependent evolution operator $U(t,-T)$.  We can impose causality by insisting that at each discrete Planck time step $\tau$, $U(\tau, - T)$ factorizes into an operator acting only in the Hilbert space of the fermionic variables in the diamond $[-T,\tau]$ and one which commutes with all those variables.  

If we now consider two diamonds whose future tips are both many Planck times from $-T$, but differ by one Planck step along the geodesic we can define
\begin{equation} e^{- i L_P H(\tau)} = U(\tau + L_P, - T) U^{-1} (\tau, - T) . \end{equation}
The geometric relation between the tangent vector we have chosen to represent by $L_0$ and the one that describes time evolution along the geodesic between the diamond tips motivates the conjecture\footnote{This conjecture echoes an obscure result in algebraic QFT that T. B. learned from Nima Lashkari.  Causal diamonds in QFT do not have density matrices, but they do have positive modular operators, which play a similar role.  If we have two nested diamonds with infinitesimally close tips then the {\it modular flow} (the unitary transformation obtained by raising the modular operator to a continuous imaginary power) indeed generates Heisenberg evolution on the operator algebra in the wedge between the two diamonds.} that $H(\tau) \propto L_0 (\tau + L_P)$. Indeed $H(\tau)$ should be a Hermitian operator in the larger Hilbert space.  Furthermore, since it should entangle the extant degrees of freedom with those that are being added, we conclude that the CFT in question cannot be free fermions.   Note that any continuous deformation of the free fermion CFT along a conformal manifold will leave the value of the central charge unchanged, so the Carlip-Solodukhin calculation will be preserved by such a deformation.

We can get more insight into what kinds of interactions are needed by trying to understand the dissipation of localized information on the holographic screen.  According to \cite{Susskind:2005js,Hayden:2007cs,Sekino:2008he} this should occur by a non-local process of ``fast scrambling".  For the one dimensional metric $e^2 (x)$ we can always introduce a new coordinate $\phi$ which measures length in Planck units
\begin{equation} \phi (y) = L_P^{-1} \int_0^y e(y^{\prime})dy^{\prime} .\end{equation}  We will normalize the length of {\it every} transverse circle to $2\pi L_P$ since the physical length is counted by the number of transverse fermion fields $\psi_n$.  
So $\phi$ goes from $0$ to $2\pi$.   The variable $\phi$ is just the angle variable in DU coordinates on the diamond. A useful intuition from the work of Carlip and Solodukhin is to think of the spatial variable $\sigma$ in the $1 + 1$ dimensional CFT as a dimensionless interval in the DU coordinate $x$, near the boundary $x = \infty$.  This is not strictly necessary, as we really only use the spectral properties of the Virasoro generator $L_0$ of the CFT to define our model.  

In terms of $\phi$, the Dirac eigenfunctions are just proportional to $e^{in\phi}$, and a cut-off keeps all modes between some wave numbers $\pm M$.  Our prescription says that we have one 1+1D fermion field on a spatial interval for each $n \in [-M,M]$.  Using the discrete Fourier transform on $Z_{2M + 1}$ we obtain linear combinations of these that represent  fields labelled by $2M+1$ discrete angles $\phi (m)$. These latticized fields are canonical free fermion fields.   The free Virasoro generators can be written as sums over either set of variables.

Beyond the total entropy, which we will return to below, another piece of physics we can try to realize in the model is the entropy deficit associated with localized particles. Let us try to find states of our system that could correspond to particles of mass $\mu$ sitting at the origin. They should have entropy deficit  $2\pi \mu R$.  Candidate states are those satisfying
\begin{equation} \psi_{\theta (m) (\sigma)} | s, \mu \rangle = 0 \hspace{1mm},\end{equation}  for $ m = 0$ mod $p$, with $p = MS_0/(\pi \mu R)$ and $S_0$ is the entropy of a single fermion field theory which we discussed above.  These states will have the proper entropy deficit.  In the free fermion theory the constraints could be removed (but would not be if we had chosen to impose them on the Fourier transforms of the fields with respect to $\sigma$) by time evolution, but information would not be homogenized over the transverse dimension, as indicated by classical field theory computations in the bulk.  This is a second indication that we need interactions.

A general class of interactions is suggested by the dual constraints of conformal invariance and transverse fast scrambling.  From each of the free fields $\psi_{\theta (m)} (\sigma)$ we can construct a $U(1)$ current $J^{\alpha}_{\theta (m)} (\sigma) $ and the interaction
\begin{equation} \delta {\cal L} = G_{mn} J^{\alpha}_{\theta (m)} (\sigma) J^{\beta}_{\theta (n)} (\sigma) \eta_{\alpha\beta}, \end{equation} defines a well defined conformal manifold of perturbations of the free fermion model for any symmetric positive definite matrix $G_{mn}$.   The eigenvalues of $L_0$ of course depend on the choice of this matrix.  For our problem we need a sequence of such choices, one for each sufficiently large sub-diamond of the diamond defined by $[-T,T]$.  At each step in time we keep all the old matrix elements of $G_{mn}$ and choose the new ones from a random distribution consistent with the constraints of symmetry and positivity.  It's clear that once $\tau$ is large, the perturbation at each time step will be small.  As far as the modular Hamiltonian is concerned, the main thing that will be affected is the probability for the previously non-existent variables, and their correlations with those that existed in the prior diamond.  It's also clear that the dynamics has no hint of locality in the transverse circle and so is likely to exhibit fast scrambling.  

By causality, we only need to talk about is time evolution within a given causal diamond, but it is convenient to have a unitary operator on the entire Hilbert space at each instant of time.  That operator should factor into a product of one that operates only on the degrees of freedom present in the diamond $[-T,\tau]$ and a unitary on its tensor complement in the full Hilbert space of the interval $[-T,T]$. We can implement this by choosing to make the coefficients $G_{mn}$ time dependent.  That is, the Hamiltonian is the sum of free fermion Hamiltonians for ALL of the fermion fields, with current current interactions added gradually as time goes on and new degrees of freedom enter the expanding diamond. Note that this means that once $\tau$ is large in Planck units, the incremental change in the Hamiltonian in each Planck time step is small.

The last remark implies that the natural time scale for any decay process, in geodesic proper time,  is of order $R$, if we adjust the coefficient relating the dimensionless $L_0$ to the Hamiltonian appropriately.   If, as we have claimed, the system is a fast scrambler, then the time scale for erasing an entropy deficit $\Delta S$ will be $C R\ {\rm ln}(\Delta S) $.   The coefficient $C$ will depend on the choice of the coefficients $G_{mn}$.

\subsection{The Cut Off Procedure}

The Cardy formula for the entropy of a CFT is given by
\begin{equation} S = 2\pi \sqrt{\frac{c}{6}(L_0 - \frac{c}{24})} . \end{equation}  Cardy and Solodukhin choose $L_0$ by finding an appropriate solution of the near horizon Liouville equation.  Of course, in a typical CFT there will not be any degeneracies and $L_0$ certainly has a discrete spectrum.  This formula represents the density of states in a particular band of eigenstates.  We can calculate the fluctuations in $L_0$ for large $c$ by integrating
\begin{equation} Z \equiv \int dL_0 e^{A\sqrt{L_0} - \beta L_0} , \end{equation}
by saddle point for large $A$ and calculating
\begin{equation} \partial^2_{\beta} ({\rm ln}\ Z)_{\beta =  1} = A \sqrt{L_0*} . \end{equation}
$A$ is fixed by fitting to the Carlip-Solodukhin (CS) formula for diamond entropy.  

This indicates a prescription for how we are to choose the two dimensional cutoff in the fermion model.  We first choose the transverse cutoff $M$ to give a value of the central charge that scales like the diamond radius for all the causal diamonds $[-T,\tau]$.  Then we fix the overall normalization by singling out a band of eigenstates of the CFTs that gives the CS result and the fluctuation formula $(\Delta K)^2 = K$ on the nose. (For more detailed discussion of the modular fluctuations, see~\cite{BZ,Banks:2022irh}.)  We emphasize that the CS result is really the first term in an asymptotic expansion in $L_P/L$, where $L$ is the size of the smallest causal diamond for which we trust the CS argument.  In the language of the analogy with hydrodynamics, we are trying to bootstrap a quantum system from knowledge of its hydrodynamic equations.  In condensed matter physics, the higher order corrections to hydrodynamic equations are replete with system dependent ambiguities, which also depend precisely on how one defines the hydrodynamic variables in terms of the microscopic degrees of freedom. The highly symmetrical structure of general relativity and quantum field theory has enabled us to make quite detailed guesses about the underlying quantum theory but one might imagine that one would need confrontation with an actual experimental system to get all of the microscopic details correct. 

We want to emphasize that the role of the two dimensional cutoff is to delineate the boundary where the hydrodynamic ansatz of Carlip and Solodukhin becomes an acceptable approximation to the dynamics of a large causal diamond in $dS_3$.  It is analogous to the choice of a block size $L$ in the derivation of hydrodynamics from the Schrodinger equation for lattice quantum systems \cite{Banks:2018aob}.  As emphasized in \cite{Banks:2018aob} there is a lot of ambiguity involved in that choice. One needs either a precise microscopic model or a comparison with experiment to tie down the details more accurately.  At the moment, we have neither.

The details of the program we have outlined will of course depend on our choice of $G_{mn} (\tau) $.   It is not clear whether this is a fundamental ambiguity or not.  We are far from exhausting all of the constraints that a sensible quantum theory of three dimensional de Sitter space must satisfy.  On the other hand, it might be that one can obtain many different consistent quantum models of $dS_3$, depending on the form of ``compact dimensions".  The latter situation seems to hold for two dimensional models of quantum gravity with linear dilaton asymptotics \cite{Callan:1992rs, Russo:1992ax}.  These arise as decoupling limits of linear dilaton black holes in four dimensional supersymmetric compactifications of Type II string theory.   There seem to be many consistent non-perturbative models of linear dilaton black holes \cite{Banks:2015mya, Banks:2020zrt}, which are plausibly related to the wide variety of consistent four dimensional string models.

\subsection{Constraints in Two Different dS Static Frames}
Let's work in conformal coordinates.  One geodesic at $\vec{y} = \vec{y}_0$, one at the origin.  The proper times are synchronized.    The cosmological horizons are just at the Euclidean distances $R$ from each of these points.   The proper time corresponding to conformal time $\eta$ is
\begin{equation}
    \tau = \int_{\eta^0}^\eta {{ds}\over  {s - \eta^0}} ,
\end{equation}
where $\eta^0$ is the place where we impose our initial condition.  At conformal time $s$ a detector on the geodesic at some point has a causal diamond that reaches out to $s \hat{n}$.  Note that the range of $s$ and $\eta^0$ could go as far back as $-\infty$.  The time-like geodesics that are static in conformal or flat coordinates all begin within a single causal diamond as shown in Fig.(\ref{fig:nestedcd}).   Thus, constraints on one geodesic have to appear as constraints on another, since constraints are by definition acausal.  They are imposed globally on the Hilbert space of the holographic screen, and make sense as an initial condition but cannot be imposed at an intermediate time without violating causality.  

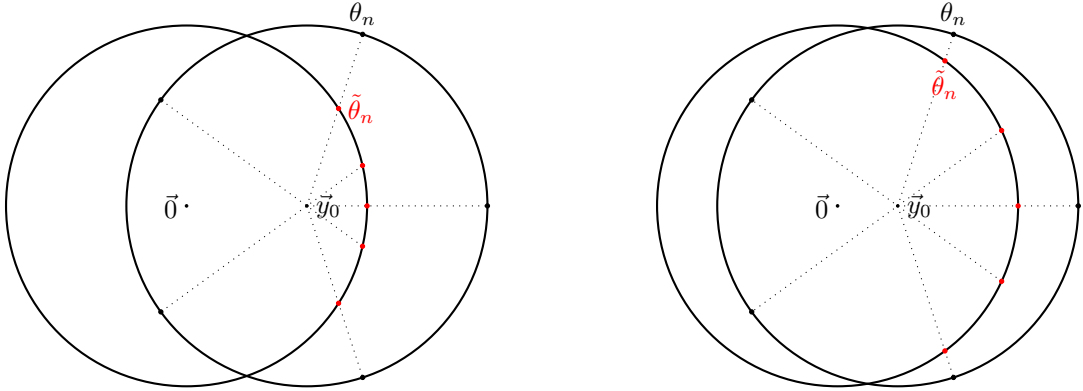
\begin{figure}
    \begin{minipage}{0.5\textwidth}
        \begin{tikzpicture}[scale=0.8]
        \draw[black, thick] (0,0) circle (3cm);
        \draw[black, thick] (2,0) circle (3cm);
        \filldraw[black] (0,0) circle (0.6pt) node[anchor=east]{$\vec{0}$};
        \filldraw[black] (2,0) circle (0.6pt) node[anchor=west]{$\vec{y}_0$};
        \draw[black,dotted] (2,0) -- (5,0);
        \draw[black,dotted] (2,0) -- (2.927,2.853);
        \draw[black,dotted] (2,0) -- (2.927,-2.853);
        \draw[black,dotted] (2.924,-0.671) -- (-0.427,1.763);
        \draw[black,dotted] (2.924,0.671) -- (-0.427,-1.763);
        \filldraw[black] (5,0) circle (1pt);
        \filldraw[black] (2.927,2.853) circle (1pt) node[anchor=south]{$\theta_n$};
        \filldraw[black] (2.927,-2.853) circle (1pt);
        \filldraw[black] (-0.427,-1.763) circle (1pt);
        \filldraw[black] (-0.427,1.763) circle (1pt);
        \filldraw[red] (3,0) circle (1pt);
        \filldraw[red] (2.924,0.671) circle (1pt);
        \filldraw[red] (2.924,-0.671) circle (1pt);
        \filldraw[red] (2.526,1.619) circle (1pt)node[anchor=west]{$\tilde{\theta}_n$};
        \filldraw[red] (2.526,-1.619) circle (1pt);
        \end{tikzpicture}
    \end{minipage}\hfill
    \begin{minipage}{0.5\textwidth}
        \centering
        \begin{tikzpicture}[scale=0.8]
        \draw[black,thick] (0,0) circle (3cm);
        \draw[black,thick] (1,0) circle (3cm);
        \filldraw[black] (0,0) circle (0.6pt) node[anchor=east]{$\vec{0}$};
        \filldraw[black] (1,0) circle (0.6pt) node[anchor=west]{$\vec{y}_0$};
        \draw[black,dotted] (1,0) -- (4,0);
        \draw[black,dotted] (1,0) -- (1.927,2.853);
        \draw[black,dotted] (1,0) -- (1.927,-2.853);
        \draw[black,dotted] (2.726,-1.254) -- (-1.427,1.763);
        \draw[black,dotted] (2.726,1.254) -- (-1.427,-1.763);
        \filldraw[black] (4,0) circle (1pt);
        \filldraw[black] (1.927,2.853) circle (1pt)node[anchor=south]{$\theta_n$};
        \filldraw[black] (1.927,-2.853) circle (1pt);
        \filldraw[black] (-1.427,-1.763) circle (1pt);
        \filldraw[black] (-1.427,1.763) circle (1pt);
        \filldraw[red] (3,0) circle (1pt);
        \filldraw[red] (2.726,1.254) circle (1pt);
        \filldraw[red] (2.726,-1.254) circle (1pt);
        \filldraw[red] (1.784,2.412) circle (1pt)node[anchor=north]{$\tilde{\theta}_n$};
        \filldraw[red] (1.784,-2.412) circle (1pt);
        \end{tikzpicture}
    \end{minipage}
    \caption{Mapping constraints on the boundary of causal diamond $D_1$ centered at $\Vec{y}_0$ onto the boundary of causal diamond $D_0$ centered at $\Vec{0}$. Each state on $D_1$ boundary outside $D_0$ is mapped to the intersection of $D_0$ boundary and a line connecting this state with $\Vec{y}_0$ (shown as a dotted line in the diagram). For states on $D_1$ boundary and inside $D_0$, extend the dotted line so it intersects with the $D_0$ boundary as shown in the diagram, then the intersection gives the mapped state. The closer $\Vec{y}_0$ is to the origin $\Vec{0}$, the more spread-out the mapped states are, and the locations of these states tell us where $\Vec{y}_0$ is.}
    \label{fig:constraintsmapping}
\end{figure}

The static geodesic at $\vec{y}_0$ leaves the causal diamond centered at the origin in a particular direction.  In conformal coordinates this happens because the diamond for $\eta > \eta^{\prime}$ shrinks in the $\vec{y}$ coordinate as $\eta^{\prime}$ is made larger, so that the geodesic at $\vec{y}_0$ is no longer causally connected to a late time detector on the geodesic at the origin.  The associated moving particle solution rapidly approaches the solution for an empty static patch, healing up a conical defect centered at $\vec{y}_0$.   We will model this behavior in the quantum theory by imposing initial conditions on the field $\psi (\theta, \sigma)$.   We will impose
$ \psi (\tilde{\theta}_n , \sigma) | S \rangle = 0 $ on the initial state of the system.   
The angles $\tilde{\theta}_n$ are determined as follows.   At very early times we consider the $Z_N$ symmetric distribution of points $\theta_n$ on the circle of radius $|\eta^0|$ surrounding ${\vec y}_0$.   We take $\eta^0$ large and negative, but finite.  Some of the $\theta_n$ are inside the circle of radius $|\eta^0|$ surrounding the origin and some are outside.  We project each $\theta_n$ to the corresponding $\tilde{\theta}_n$ by drawing the line connecting $\theta_n$ to $\vec{y}_0$.  $\tilde{\theta}_n$  is the point where that line intersects the circle surrounding the origin as shown in Fig.(\ref{fig:constraintsmapping}).

Note that these constraints are imposed on the full set of microscopic degrees of freedom of the theory.   The dynamical evolution of the system does not couple these degrees of freedom together immediately.  At each proper Planck time step, a finite set of angular momentum modes interact via fast scrambling interactions of the maximally transversely localized $U(1)$ currents that we can form from them.  At early times these currents do not have much overlap with the highly localized fields at individual $\theta_n$, when the de Sitter radius is large in Planck units.  The rest of the individual angular momentum modes evolve as free fermion fields, all obeying the same linear equation of motion.  Thus, their contribution to the constraint equation is undisturbed by time evolution.   
As the proper time approaches the dS Hubble time and beyond, all the modes participate in the fast scrambling interaction, and the constraints are removed.   It's also clear that the further $\vec{y}_0$ is from the origin, the constrained state is more of a distortion of the homogeneous dS equilibrium state.  Since the fast scrambling interactions immediately transfer information between points that are separated by arbitrary distances on the lattice, these will homogenize rapidly.   

Much work remains to be done to find a more quantitative match between lifetimes and entropy deficits for particles in $dS_3$ as defined by the boundary quantum system, when compared to the semi-classical calculations based on the KG equation. The general qualitative behavior of our explicit quantum system clearly mirrors the semi-classical physics we have explored.  In the Appendix we begin a more quantitative exploration by computing the overlap diamonds for pairs of intervals along different geodesics.

\section{Conclusions}

We have argued that the semi-classical physics of particles propagating in $dS_3$, is compatible with a quantum theory based on the idea of static patch holography \cite{tbwf}.  The appropriate quantum framework is a Hilbert bundle over the space of time-like trajectories in $dS_3$ with a finite dimensional Hilbert space over each fiber. This is a more refined version of the HST formalism introduced by two of the authors. The dS group $SO(1,3)$ acts on this bundle, with a typical transformation mapping one fiber into another.   Within each fiber only the $SO(2) \times R$ subgroup leaving a particular geodesic invariant, acts.  Dynamics within each fiber is provided by time dependent Hamiltonians adapted to either past or future oriented nests of causal diamonds.  The future oriented nest provides us with a mathematical description of how the whole $dS_3$ universe evolves, given initial conditions on the past boundary of the static patch.  The past oriented nest is useful for understanding what a detector traveling along the geodesic can probe after a given time.

The symmetry action guarantees that the dynamics along each geodesic is identical.  Similarly, time reversal relates the Hamiltonians for the backward and forward oriented nests of diamonds.  In both cases, the complication lies in mapping a particular initial state along one sequence of nested proper time intervals on a particular geodesic to its corresponding state in another nesting or another geodesic.  We have made a start on understanding this for the constrained states corresponding to single particles traveling on geodesics, whose relative velocity in flat coordinates is zero.  

Perhaps the most disturbing consequence of our picture is the complete failure of quantum field theory in curved space-time to give any sort of approximation to our picture.
One can view this as a drastic version of the Cohen-Kaplan-Nelson bound \cite{CKN,Banks:2019arz,Blinov:2021fzl, Banks:2020tox} on the validity of QFT.  Most states of QFT in $dS_3$ are simply not realizable once the classical backreaction of matter on geometry is taken into account.  This is dramatically different from the situation in $AdS_3$.   Although the classical analysis of particles coupled to $3D$ gravity leads to similar conclusions for either sign of the cosmological constant, the Bekenstein-Hawking principle tells us that that description is missing a huge number of quantum states compatible with asymptotically $AdS_3$ geometry.  BTZ black holes have large entropy.  In $dS_3$ the CEP identifies missing states on the cosmological horizon but not in the bulk interior.  Thus, there seems to be a complete model of a small finite number of particles, with restricted phase space, interacting only with gravity.  Note that this is a self consistent picture.  In $2+1$ dimensions, gravitational interactions do not lead to processes with particle-anti-particle production.

Asymptotically flat 3 dimensional space, to the extent such a concept exists in real models of quantum gravity, would appear to be in the same class as $dS_3$.   The argument that CFT correlators converge to S matrix elements fails in three dimensions, because the S-matrix does not exist, and any possible notion of scattering amplitudes for finite numbers of particles contains bizarre constraints on Mandelstam invariants.  This is consistent with the empirical fact that every model of $AdS_3/CFT$ with an $R_{AdS}/l_S \rightarrow\infty$ limit has two or more compact dimensions whose radius (inverse mass of the lightest Kaluza-Klein state) goes to infinity linearly with $R_{AdS}$.    Although it is conceivable that a model in asymptotically locally flat space could be defined by taking the $N\rightarrow\infty$ limit of our model of $dS_3$, we make no such claim at this time.

The failure of quantum field theory in curved space time to adequately approximate our model deserves further study.  A possible starting point is a model of charged particles coupled to electromagnetism and gravity.  Such a model has local vertices, which lead to particle-anti-particle production.  If the particles are sufficiently light, then the geometric back reaction on a single such process is minimal.  It's clear however that multiparticle states with any finite energy per particle are forbidden by gravitational constraints whenever the particle number is too high.   Thus, it would appear that neither the constructions of the present paper, nor conventional field theory descriptions can model this situation.  It might lie in the class of low energy field theory models that do not arise as limits of models of quantum gravity.  

Another feature of our analysis that we found surprising is that so far we do not have a time independent Hamiltonian description of our static patch model in static time.  We've instead exploited a plausible connection between modular flow and Planck step geodesic evolution in the DU coordinates of \cite{JV}.  In QFT, the Heisenberg evolution of operators in a small wedge between closely spaced nested diamonds can be related to modular flow, but is also generated by the action of the global Hamiltonian operator, restricted to the subalgebra of the diamond.  One would have hoped that something similar happened in $dS$ space, but so far we have not found the global description.  

\begin{center}
{\bf Acknowledgments }\\
We thank Patrick Draper for collaboration in the early stages of this work and for some ideas contributed to section \ref{sec:IIIB}. The work of T. B. and S. A is supported in part by the DOE under Grant DE-SC0010008. The work of W. F. is supported by the NSF under Grant PHY-1914679.
\end{center}

\appendix
\addappheadtotoc
\appendixpage

\begin{figure}
    \centering
    \begin{minipage}{0.5\textwidth}
        \centering
        \begin{tikzpicture}[scale=0.8]
        \filldraw[blue!8!white!] (0.75,0) circle (1.25cm);
        \draw[black, thick] (0,0) circle (2cm);
        \draw[black, thick] (2.5,0) circle (3cm);
        \draw[blue] (0.75,0) circle (1.25cm);
        \filldraw[black] (0,0) circle (0.6pt) node[anchor=east]{$\vec{0}$};
        \filldraw[black] (2.5,0) circle (0.6pt) node[anchor=west]{$\vec{y}_0$};
        \filldraw[blue] (0.75,0) circle (0.6pt);
        \draw[blue] (0.75,0)--(2,0) node[anchor=north]{$R_{overlap}$};
        \end{tikzpicture}
    \end{minipage}\hfill
    \begin{minipage}{0.5\textwidth}
        \centering
        \begin{tikzpicture}[scale=0.8]
        \filldraw[blue!8!white!] (-0.5,0)--(0.75,1.25)--(2,0)--(0.75,-1.25)--(-0.5,0);
        \draw [dashed] (2,0) arc[x radius=2, y radius=0.26, start angle=0, end angle=180];
        \draw [dashed] (5.5,0) arc[x radius=3, y radius=0.39, start angle=0, end angle=180];
        \draw (5.5,0) arc[x radius=3, y radius=0.39, start angle=0, end angle=-180];
        \draw (2,0) arc[x radius=2, y radius=0.26, start angle=0, end angle=-180];
        \draw[blue, thick] (2,0) arc[x radius=1.25, y radius=0.16, start angle=0, end angle=-180];
        \draw[blue, thick, dashed] (2,0) arc[x radius=1.25, y radius=0.12, start angle=0, end angle=180];
        \draw(3,-1) arc[x radius=3, y radius=0.39, start angle=0, end angle=-180];
        \draw[dashed] (3,-1) arc[x radius=3, y radius=0.39, start angle=0, end angle=180];
        \draw[black, thick] (-3,-1)--(0,2)--(3,-1)--(0,-4)--(-3,-1);
        \draw[black, thin,dotted] (0,2)--(0,-4);
        \draw[black, thick] (-0.5,0)--(2.5,3)--(5.5,0)--(2.5,-3)--(-0.5,0);
        \draw[black, thin,dotted] (2.5,3)--(2.5,-3);
        \draw[black,thick] (0.75,-1.25)--(2,0);
        \filldraw[black] (0,-1) circle (0.6pt) node[anchor=east]{$\vec{0}$};
        \filldraw[black] (2.5,0) circle (0.6pt) node[anchor=west]{$\vec{y}_0$};
        \filldraw[blue] (0.75,0) circle (0.6pt);
        \filldraw (0,-4) circle (0pt) node[anchor=north]{$t_1$};
        \filldraw (0,2) circle (0pt) node[anchor=south]{$t_2$};
        \filldraw (2.5,-3) circle (0pt) node[anchor=north]{$t'_1$};
        \filldraw (2.5,3) circle (0pt) node[anchor=south]{$t'_2$};
        \end{tikzpicture}
    \end{minipage}
    \caption{An example top view (left) of intersecting spatial slices of two diamonds (right). The maximal overlap has radius $R_{overlap}$. This example corresponds to the case in Fig. \ref{fig:R-y0<yy'}, the particle at $\Vec{0}$ is enclosed in the maximal overlap, leading to a deficit angle $\delta$.}
    \label{fig:topdown}
\end{figure}
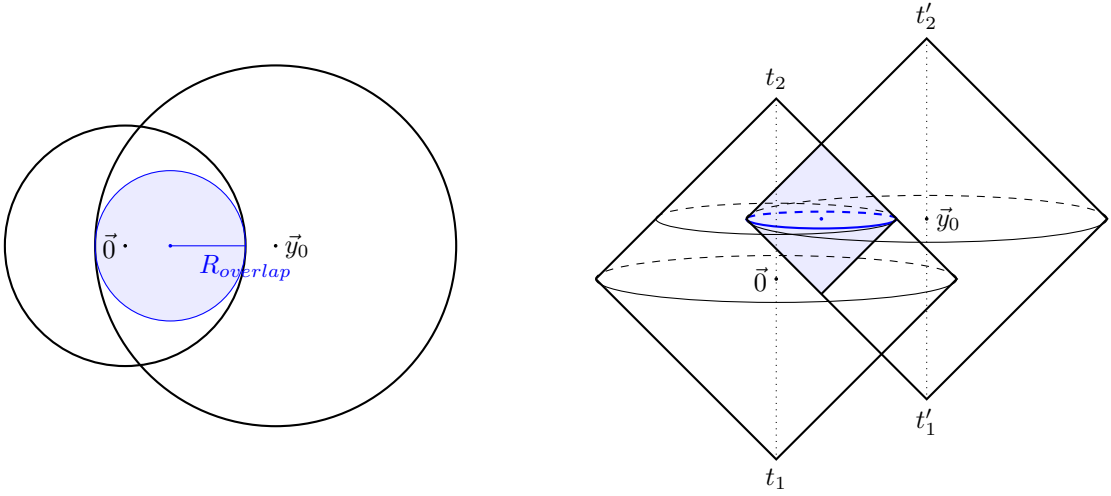

\hspace{5mm} The entropy of the overlap of two causal diamonds is computed in this section. A causal diamond is a region accessible to a detector on an interval of a time-like trajectory. The intersection between two diamonds in $2+1$ dimensions has a  boundary that ends, in general, in two lines rather than two points. One looks at all time-like geodesics connecting a point on the past line with a point on the future line and chooses the one whose diamond has the largest area. Fig. \ref{fig:topdown} gives an example of an intersection of two diamonds.\footnote{We emphasize that time coordinates $t_1,t_2, t_1',t_2'$ labelled in all figures in this Appendix are flat time coordinates,  and constant flat time slices are curved in the conformal diagram Fig. \ref{fig:nestedcd}. In the conformal diagram, a causal diamond along a timelike geodesic $\vec{y}_0$ has a rectangular side view, and it is stretched to a square diamond in Fig. \ref{fig:topdown}-\ref{fig:R-y0<yy'} where the horizontal and vertical directions are in flat space and conformal time coordinates, respectively.} In 2+1 dimensions, the area reduces to the circumference, thus to determine the entropy we want to find the overlap radius. Start with the flat-slicing metric in de Sitter space Eq.(\ref{eq:flatslicing}), where $R$ is de Sitter radius. A general solution for time-like geodesics is given by Eq.(\ref{eq:time-likesolution}), and constant $\Vec{y}=\Vec{y}_0$ is the trial solution when $\Vec{v}_0=0$. For null geodesics, vanishing spacetime intervals give the following general solution \begin{equation}\label{eq:nullgeodesic}
    \Vec{y}_{null}(t) = \Vec{u}+\Vec{v}Re^{-t/R} \hspace{1mm},
\end{equation} where $\vec{u}$ and unit vector $\vec{v}$ are parameters to be determined. Note that in a causal diamond, horizontal time slices correspond to constant conformal times $\eta$, which is related to the flat time $t$ by $d\eta/dt = e^{-t/R}$. Constant time flat slices are yellow curves in Fig. \ref{fig:nestedcd}.

Now consider a causal diamond along time-like geodesic $\Vec{y}=0$. Let $(t_1, t_2)$ be a time interval along $\Vec{y}=0$, and the past and future null boundaries of the causal diamond cross $\Vec{y}=0$ at $t_1$ and $t_2$, respectively. Along with the intersection of the past and future boundaries of the causal diamond, we obtain three equations to determine parameters $\Vec{u}$ and $\Vec{v}$,
\begin{equation}\label{eq:vector}
    \begin{split}
        \Vec{u}_1+\Vec{v}_1Re^{-t_1/R} &= 0 \hspace{1mm}, \\
        \Vec{u}_2+\Vec{v}_2Re^{-t_2/R} &= 0  \hspace{1mm}, \\
        \Vec{u}_1+\Vec{v}_1Re^{-t_0/R} &=
        \Vec{u}_2+\Vec{v}_2Re^{-t_0/R} \hspace{1mm},
    \end{split}
\end{equation}
where $t_0$ is the flat time  of the bifurcation surface. In terms of conformal time $\eta=-Re^{-t/R}$, we have $\eta_0=(\eta_1+\eta_2)/2$.
Solving Eq. (\ref{eq:vector}) leaves us one parameter, which is the direction of a free unit vector $\Vec{v}_1$. Express $\Vec{v}_2, \Vec{u}_1, \Vec{u}_2$ in terms of unit vector $\vec{v}_1$, we find the boundaries of the causal diamond to be
\begin{equation}
    \begin{split}
        \vec{y}_1(t) & = \vec{v}_1R\Big(e^{-t/R}-e^{-t_1/R}\Big) \hspace{1mm}, \\
        \vec{y}_2(t) & = -\vec{v}_1R\Big(e^{-t/R}-e^{-t_2/R}\Big) \hspace{1mm}.
    \end{split}
\end{equation}

Let $e^{-t_0/R}=(e^{-t_1/R}+e^{-t_2/R})/2$ to solve for the intersection $\vec{y}_{int}$ of the past and future null boundaries, and we find the radius of the causal diamond to be
\begin{equation}\label{eq:yint}
    \begin{split}
        y_{int} = \frac{R}{2}\Big(e^{-t_1/R} - e^{-t_2/R}\Big) \hspace{1mm}.
    \end{split}
\end{equation}
Then the area of a causal diamond is given by the circumference of the bifurcation surface, which is the two-dimensional surface centred at $\vec{y}=0$ with boundary $\vec{y}_{int}$. The area of the causal diamond is
\begin{equation}
    \begin{split}
        A = \Omega y_{int}  = \frac{1}{2}\Omega R\Big(e^{-t_1/R} - e^{-t_2/R}\Big) \hspace{1mm},
    \end{split}
\end{equation}
where $\Omega = 2\pi - \delta$ with $\delta = 8\pi G_N m$ \cite{dj} being the deficit angle due to the presence of a massive particle with mass $m$.

Now let us consider two maximal causal diamonds with dS radius $R$ whose time runs from $-\infty$ to $\infty$, centred along two time-like trajectories at $\Vec{y}=0$ and $\Vec{y}=\Vec{y}_0$. Two time-intervals $(t_1, t_2)$ and $(t_1', t_2')$ along the two trajectories give two smaller causal diamonds. When the two diamonds overlap, there exists a maximal causal diamond within the overlapping region and we are interested in determining the entropy of this maximal overlap, the result depends on $y_0, t_1, t_2, t_1', t_2'$ and we will discuss by cases. 

\vspace{3mm}
\noindent\textbf{\Large{A. When $y_0>2R$}}
\vspace{2mm}

The two maximal causal diamonds do not overlap, then no matter how large $(t_1, t_2)$ and $(t_1', t_2')$ are, there is no overlap. This case corresponds to when the causal diamond centered at $\Vec{y}_0$ is completely unobservable viewed by an observer along $\Vec{y}=0$. See Fig. \ref{fig:>2R} for a front view of the case.
\vspace{2mm}
\begin{figure}[ht]
    \centering
    \begin{minipage}{0.48\textwidth}
        \centering
        \hspace{5mm}
        \begin{tikzpicture}
        \draw[black,thick] (0,0) -- (2,2) -- (4,0) -- (2,-2) -- (0,0);
        \draw[black, dotted] (2,-2) -- (2,2);
        \draw[red,ultra thin] (1.1,0.4) -- (2,1.3) -- (2.9,0.4) -- (2,-0.5) -- (1.1,0.4);
        \filldraw[black] (2,-2) circle (0pt) node[anchor=north]{$\Vec{0}$};
        \filldraw[red] (2,-0.5) circle (0pt) node[anchor=north]{$t_1$};
        \filldraw[red] (2,1.3) circle (0pt) node[anchor=south]{$t_2$};

        \draw[black, thick] (4.5,0) -- (6.5,2) -- (8.5,0) -- (6.5,-2) -- (4.5,0);
        \draw[black, dotted] (6.5,-2) -- (6.5,2);
        \draw[red,ultra thin] (5.1,0.2) -- (6.5,1.6) -- (7.9,0.2) -- (6.5,-1.2) --(5.1,0.2);
        \filldraw[black] (6.5,-2) circle (0pt) node[anchor=north]{$\vec{y}_0$};
        \filldraw[red] (6.5,-1.2) circle (0pt) node[anchor=south]{$t'_1$};
        \filldraw[red] (6.5,1.6) circle (0pt) node[anchor=north]{$t'_2$};
        \end{tikzpicture}
        \caption{When $y_0>2R$, there is no overlap between causal diamonds $(t_1,t_2)$ and $(t_1',t_2')$.}
        \label{fig:>2R}
    \end{minipage}\hfill
    \begin{minipage}{0.48\textwidth}
        \centering
        \hspace{3mm}
        \begin{tikzpicture}
        \draw[black, thick] (0,0) -- (2,2) -- (4,0) -- (2,-2) -- (0,0);
        \draw[black, dotted] (2,-2) -- (2,2);
        \draw[line width=0.01,red] (1,0) -- (2,1) -- (3,0) -- (2,-1) -- (1,0);
        \filldraw[black] (2,-2) circle (0pt) node[anchor=north]{$\Vec{0}$};
        \filldraw[red] (2,-1) circle (0pt) node[anchor=north]{$t_a$};
        \filldraw[red] (2,1) circle (0pt) node[anchor=south]{$t_b$};
        
        \draw[black, thick] (3,0) -- (5,2) -- (7,0) -- (5,-2) -- (3,0);
        \draw[black, dotted] (5,-2) -- (5,2);
        \filldraw[black] (5,-2) circle (0pt) node[anchor=north]{$\vec{y}_0$};
        \end{tikzpicture}
        \caption{When $R<y_0<2R$ and causal diamonds centered at $\Vec{y}_0$ and $(t_a,t_b)$ meet.}
        \label{fig:R-2R}
    \end{minipage}
\end{figure}

\begin{figure}[ht]
    \centering
    \begin{minipage}{0.48\textwidth}
        \centering
        \begin{tikzpicture}
        \draw[black, thick] (0,0) -- (2,2) -- (4,0) -- (2,-2) -- (0,0);
        \draw[black, dotted] (2,-2) -- (2,2);
        \draw[line width=0.01,red] (0.3,0.1) -- (2,1.8) -- (3.7,0.1) -- (2,-1.6) -- (0.3,0.1);
        \filldraw[black] (2,-2) circle (0pt) node[anchor=north]{$\Vec{0}$};
        \filldraw[red] (2,-1.6) circle (0pt) node[anchor=south]{$t_1$};
        \filldraw[red] (2,1.8) circle (0pt) node[anchor=north]{$t_2$};
        
        \draw[black, thick] (3,0) -- (5,2) -- (7,0) -- (5,-2) -- (3,0);
        \draw[line width=0.01,red] (3.7,0.1) -- (5,1.4) -- (6.3,0.1) -- (5,-1.2) -- (3.7,0.1);
        \draw[black, dotted] (5,-2) -- (5,2);
        \filldraw[black] (5,-2) circle (0pt) node[anchor=north]{$\vec{y}_0$};
        \filldraw[red] (5,-1.2) circle (0pt) node[anchor=south]{$t'_1$};
        \filldraw[red] (5,1.4) circle (0pt) node[anchor=north]{$t'_2$};
        \end{tikzpicture}
        \caption{When $R<y_0<2R$ and causal diamonds $(t_1,t_2)$ and $(t_1',t_2')$ meet.}
        \label{fig:R-2R-1}
    \end{minipage}\hfill
    \begin{minipage}{0.48\textwidth}
        \centering
        \begin{tikzpicture}
        \draw[black, thick] (0,0) -- (2,2) -- (4,0) -- (2,-2) -- (0,0);
        \draw[black, dotted] (2,-2) -- (2,2);
        \draw[line width=0.01,red] (0.3,0.1) -- (2,1.8) -- (3.7,0.1) -- (2,-1.6) -- (0.3,0.1);
        \filldraw[black] (2,-2) circle (0pt) node[anchor=north]{$\Vec{0}$};
        \filldraw[red] (2,-1.6) circle (0pt) node[anchor=south]{$t_1$};
        \filldraw[red] (2,1.8) circle (0pt) node[anchor=north]{$t_2$};
        
        \draw[black, thick] (3,0) -- (5,2) -- (7,0) -- (5,-2) -- (3,0);
        \draw[line width=0.01,red] (3.15,-0.05) -- (5,1.8) -- (6.85,-0.05) -- (5,-1.9) -- (3.15,-0.05);
        \draw[black, dotted] (5,-2) -- (5,2);
        \draw[blue] (3.7,0.1) -- (3.3,0.1);
        \filldraw[blue] (3.7,0.3) circle (0pt) node[anchor=west]{$2R_{overlap}$};
        \filldraw[black] (5,-2) circle (0pt) node[anchor=north]{$\vec{y}_0$};
        \filldraw[red] (5,-1.9) circle (0pt) node[anchor=south]{$t'_1$};
        \filldraw[red] (5,1.8) circle (0pt) node[anchor=north]{$t'_2$};
        \end{tikzpicture}
        \caption{When $R<y_0<2R$ and causal diamonds $(t_1,t_2)$ and $(t_1',t_2')$ overlap.}
        \label{fig:R-2R-2}
    \end{minipage}
\end{figure}

\noindent\textbf{\Large{B. When $R<y_0<2R$}}
\vspace{2mm}

This case corresponds to when the particle along $\Vec{y}_0$ trajectory is outside the observable universe, but part of its causal diamond overlaps with the causal diamond centered at $\Vec{y}=0$. $(t_1, t_2)$ and $(t_1', t_2')$ need to satisfy conditions to allow overlap (which we will show below), then the entropy is given by $2\pi R_{overlap}$ since neither time-like trajectory passes through the overlapping causal diamond, thus the deficit angle is $0$. The intersection of minor null surfaces of small causal diamond with $(t_1, t_2)$ is given by Eq.(\ref{eq:yint}), which is the radius of the red diamond in Fig. \ref{fig:R-2R}. As shown in the Figure, any time interval with larger $t_1$ or smaller $t_2$ than such $(t_1, t_2)$ in Fig. \ref{fig:R-2R} gives no overlap, i.e. overlap exists when
\begin{equation}\label{eq:conditionontimetoallowoverlap}
    t_1<t_a=-R\ln \frac{y_0}{R} \hspace{5mm} \text{and} \hspace{5mm} t_2>t_b= - R\ln \bigg(2-\frac{y_0}{R}\bigg) \hspace{1mm}.
\end{equation}
Consider when $t_1, t_2$ satisfy the above conditions, Fig. \ref{fig:R-2R-1} gives the critical case when there is no overlap, which depends on $(t_1',t_2')$. Similarly let $y'_{int}$ be the radius of minor causal diamond with time $(t_1',t_2')$, then
\begin{equation}
    y'_{int} = \frac{R}{2}\Big(e^{-t'_1/R} - e^{-t'_2/R}\Big) \hspace{1mm}.
\end{equation}

Solve $y_{int}+y'_{int}=y_0$ for conditions on $t_1', t_2'$ to allow overlap, where conformal time $\eta_1'+\eta_2'=\eta_1+\eta_2$. We find that overlap exists only if
\begin{equation}\label{eq:t'cond}
    \begin{split}
         t_1' &< -R\ln \bigg(e^{-t_2/R}+\frac{y_0}{R}\bigg) \hspace{1mm},\\
        \text{and} \hspace{5mm} t'_2 &> - R\ln \bigg(e^{-t_1/R}-\frac{y_0}{R}\bigg) \hspace{1mm}.
    \end{split}
\end{equation}

Now consider when all conditions are met, i.e.
when Eq. (\ref{eq:conditionontimetoallowoverlap}) and Eq. (\ref{eq:t'cond}) are satisfied. Then overlap exists, as shown in Fig. \ref{fig:R-2R-2}, and $R_{overlap}$ is a function of $y_0, t_1, t_2, t_1', t_2'$ which is given by
\begin{equation}\label{eq:roverlap}
    \begin{split}
        R_{overlap} &= \frac{1}{2}\Big[y_{int}+y'_{int}\Big((\eta_1+\eta_2)/2\Big)-y_0\Big] \hspace{1mm}.
    \end{split}
\end{equation}
Here $y'_{int}\Big((\eta_1+\eta_2)/2\Big)$ is the radius of the $(t'_1,t'_2)$ causal diamond evaluated at flat time $t$ whose conformal time is $(\eta_1+\eta_2)/2$.
\begin{equation}
    y'_{int}\Big((\eta_1+\eta_2)/2\Big) = \begin{cases} \frac{R}{2}\Big(e^{-t_1/R} +e^{-t_2/R} - 2e^{-t_2'/R}\Big) \hspace{8mm} &\text{if} \hspace{3mm} \eta_1'+\eta_2'<\eta_1+\eta_2 \hspace{1mm}, \\
    \frac{R}{2}\Big(2e^{-t_1'/R} - e^{-t_1/R}-e^{-t_2/R}\Big) \hspace{8mm} &\text{if} \hspace{3mm} \eta_1'+\eta_2'\geq \eta_1+\eta_2 \hspace{1mm}.
    \end{cases}
\end{equation}
Note that when $\eta_1+\eta_2=\eta_1'+\eta_2'$, the radius $y_{int}'\Big((\eta_1+\eta_2)/2\Big)=y_{int}'$.

\vspace{3mm}
\noindent\textbf{\Large{C. When $y_0<R$}}
\vspace{2mm}

There are 3 cases: 

1. when neither trajectory is enclosed by the overlap, the deficit angle is zero; 

2. when overlapping causal diamond encloses one trajectory, one particle contributes to the deficit angle, either $\delta$ or $\delta'$ depending on which particle is enclosed; 

3. when the overlap encloses both trajectories, both particles contribute, i.e. the deficit angle is $\delta+\delta'$.

\begin{figure}[ht]
    \centering
    \begin{minipage}{0.48\textwidth}
        \centering
        \begin{tikzpicture}
        \draw[black, thick] (0,0) -- (2,2) -- (4,0) -- (2,-2) -- (0,0);
        \draw[black, dotted] (2,-2) -- (2,2);
        \draw[line width=0.01,red] (1.41,0) -- (2,0.59) -- (2.59,0) -- (2,-0.59) -- (1.41,0);
        \filldraw[black] (2,-2) circle (0pt) node[anchor=north]{$\Vec{0}$};
        \filldraw[red] (2,-0.6) circle (0pt) node[anchor=north]{$t_a$};
        \filldraw[red] (2,0.6) circle (0pt) node[anchor=south]{$t_b$};
        
        \draw[black, thick] (1.4,0) -- (3.4,2) -- (5.4,0) -- (3.4,-2) -- (1.4,0);
        \draw[black, dotted] (3.4,-2) -- (3.4,2);
        \filldraw[black] (3.4,-2) circle (0pt) node[anchor=north]{$\vec{y}_0$};
        \end{tikzpicture}
        \caption{When $y_0<R$, there is no overlap if $t_1>t_b$ or $t_2<t_a$.}
        \label{fig:R}
    \end{minipage}\hfill
    \begin{minipage}{0.48\textwidth}
        \centering
        \begin{tikzpicture}
        \draw[black, thick] (0,0) -- (2,2) -- (4,0) -- (2,-2) -- (0,0);
        \draw[black, dotted] (2,-2) -- (2,2);
        \draw[line width=0.01,red] (0.8,0.1) -- (2,1.3) -- (3.2,0.1) -- (2,-1.1) -- (0.8,0.1);
        \filldraw[black] (2,-2) circle (0pt) node[anchor=north]{$\Vec{0}$};
        \filldraw[red] (2,-1.1) circle (0pt) node[anchor=north]{$t_1$};
        \filldraw[red] (2,1.3) circle (0pt) node[anchor=south]{$t_2$};
        
        \draw[black, thick] (1.4,0) -- (3.4,2) -- (5.4,0) -- (3.4,-2) -- (1.4,0);
        \draw[line width=0.01,red] (2.05,-0.3) -- (3.4,1.05) -- (4.75,-0.3) -- (3.4,-1.65) -- (2.05,-0.3);
        \draw[black, dotted] (3.4,-2) -- (3.4,2);
        \draw[blue] (3.2,0.1) -- (2.45,0.1);
        \filldraw[blue] (2.9,0.1) circle (0pt) node[anchor=north]{$2R_{overlap}$};
        \filldraw[black] (3.4,-2) circle (0pt) node[anchor=north]{$\vec{y}_0$};
        \filldraw[red] (3.4,-1.6) circle (0pt) node[anchor=south]{$t'_1$};
        \filldraw[red] (3.4,1.05) circle (0pt) node[anchor=south]{$t'_2$};
        \end{tikzpicture}
        \caption{When $y_0<R$ and neither trajectory is enclosed in the overlap.}
        \label{fig:R-yy'<y0}
    \end{minipage}
\end{figure}
Determine range for $t_1, t_2$ where $\eta_1+\eta_2=-R$ by symmetry about $\eta=-R$ as shown in Fig. \ref{fig:R}, we find that overlap exists when
\begin{equation}
    t_1<t_b= -R\ln\frac{y_0}{R} \hspace{5mm} \text{and} \hspace{5mm} t_2>t_a=-R\ln\bigg(2-\frac{y_0}{R}\bigg) \hspace{1mm}.
\end{equation}
Now consider conditions on $t_1', t_2'$ to allow overlap. 

\vspace{2mm}
\noindent\textbf{\large{C.1 When $y_{int}\leq y_0$}}
\vspace{2mm}

The conditions on $t_1', t_2'$ to allow overlap is determined by solving $y_{int}+y'_{int}=y_0$ with $\eta_1+\eta_2=\eta_1'+\eta_2'$ just like in case 2, and we get the same result, that overlap exists only if $(t_1', t_2')$ satisfy Eq.(\ref{eq:t'cond}).

\vspace{2mm}
\textbf{C.1.1 When $y'_{int} \leq y_0$}
\vspace{2mm}

See Fig. \ref{fig:R-yy'<y0}, then results from case 2 hold, that the entropy is given by $2\pi R_{overlap}$. 

\vspace{2mm}
\textbf{C.1.2 When $y'_{int}>y_0$} \label{subsec:3a2}
\vspace{2mm}

See Fig. \ref{fig:R-y<y0<y'}, here $y'_{int}<R$ holds automatically because $t_1'>-\infty$ and $t_2'<\infty$, whether or not there is a deficit angle depends on the following.

\textbf{a}. If $\eta_1+\eta_2=\eta_1'+\eta_2'$, the entropy of overlapping causal diamond is $(2\pi-\delta)R_{overlap}$ where $\delta$ is the deficit angle due to particle at $y=0$. 

\textbf{b}. If $\eta_1+\eta_2\neq \eta_1'+\eta_2'$ and $y_{int}'((\eta_1+\eta_2)/2)\leq y_0$, the entropy is $2\pi R_{overlap}$. 

\textbf{c}. If $\eta_1+\eta_2\neq \eta_1'+\eta_2'$ and $y_{int}'((\eta_1+\eta_2)/2)> y_0$, the entropy is $(2\pi-\delta)R_{overlap}$ where $\delta$ is the deficit angle due to particle at $y=0$.

\begin{figure}[ht]
    \centering
    \begin{minipage}{0.48\textwidth}
        \centering
        \begin{tikzpicture}
        \draw[black, thick] (0,0) -- (2,2) -- (4,0) -- (2,-2) -- (0,0);
        \draw[black, dotted] (2,-2) -- (2,2);
        \draw[line width=0.01,red] (0.8,0.1) -- (2,1.3) -- (3.2,0.1) -- (2,-1.1) -- (0.8,0.1);
        \filldraw[black] (2,-2) circle (0pt) node[anchor=north]{$\Vec{0}$};
        \filldraw[red] (2,-1.1) circle (0pt) node[anchor=north]{$t_1$};
        \filldraw[red] (2,1.3) circle (0pt) node[anchor=south]{$t_2$};
        
        \draw[black, thick] (1.4,0) -- (3.4,2) -- (5.4,0) -- (3.4,-2) -- (1.4,0);
        \draw[line width=0.01,red] (1.8,-0.2) -- (3.4,1.4) -- (5,-0.2) -- (3.4,-1.8) -- (1.8,-0.2);
        \draw[black, dotted] (3.4,-2) -- (3.4,2);
        \draw[blue] (3.2,0.1) -- (2.1,0.1);
        \filldraw[blue] (2.8,0.1) circle (0pt) node[anchor=north]{$2R_{overlap}$};
        \filldraw[black] (3.4,-2) circle (0pt) node[anchor=north]{$\vec{y}_0$};
        \filldraw[red] (3.4,-1.8) circle (0pt) node[anchor=south]{$t'_1$};
        \filldraw[red] (3.4,1.4) circle (0pt) node[anchor=north]{$t'_2$};
        \end{tikzpicture}
        \caption{Whether the maximal overlap encloses the trajectory is conditional.}
        \label{fig:R-y<y0<y'}
    \end{minipage}\hfill
    \begin{minipage}{0.48\textwidth}
        \centering
        \begin{tikzpicture}
        \draw[black, thick] (0,0) -- (2,2) -- (4,0) -- (2,-2) -- (0,0);
        \draw[black, dotted] (2,-2) -- (2,2);
        \draw[line width=0.01,red] (0.4,-0.1) -- (2,1.5) -- (3.6,-0.1) -- (2,-1.7) -- (0.4,-0.1);
        \draw[line width=0.01,red] (3.2,-0.1) -- (3.4,0.1) -- (3.6,-0.1) -- (3.4,-0.3) -- (3.2,-0.1);
        \filldraw[black] (2,-2) circle (0pt) node[anchor=north]{$\Vec{0}$};
        \filldraw[red] (2,-1.7) circle (0pt) node[anchor=south]{$t_1$};
        \filldraw[red] (2,1.4) circle (0pt) node[anchor=north]{$t_2$};
        \filldraw[red] (3.4,-0.3) circle (0pt) node[anchor=north]{$t_a'$};
        \filldraw[red] (3.4,0.1) circle (0pt) node[anchor=south]{$t_b'$};
        
        \draw[black, thick] (1.4,0) -- (3.4,2) -- (5.4,0) -- (3.4,-2) -- (1.4,0);
        \draw[black, dotted] (3.4,-2) -- (3.4,2);
        \filldraw[black] (3.4,-2) circle (0pt) node[anchor=north]{$\vec{y}_0$};
        \end{tikzpicture}
        \caption{When $y_{int}>y_0$, there is no overlap when $t_1'>t_b'$ or $t_2'<t_a'$.}
        \label{fig:R-y0<y}
    \end{minipage}
\end{figure}

\begin{figure}[ht]
    \centering
    \begin{minipage}{0.48\textwidth}
        \centering
        \begin{tikzpicture}
        \draw[black, thick] (0,0) -- (2,2) -- (4,0) -- (2,-2) -- (0,0);
        \draw[black, dotted] (2,-2) -- (2,2);
        \draw[line width=0.01,red] (0.4,-0.1) -- (2,1.5) -- (3.6,-0.1) -- (2,-1.7) -- (0.4,-0.1);
        \filldraw[black] (2,-2) circle (0pt) node[anchor=north]{$\Vec{0}$};
        \filldraw[red] (2,-1.7) circle (0pt) node[anchor=south]{$t_1$};
        \filldraw[red] (2,1.5) circle (0pt) node[anchor=north]{$t_2$};
        
        \draw[black, thick] (1.4,0) -- (3.4,2) -- (5.4,0) -- (3.4,-2) -- (1.4,0);
        \draw[line width=0.01,red] (2.2,0.2) -- (3.4,1.4) -- (4.6,0.2) -- (3.4,-1) -- (2.2,0.2);
        \draw[black, dotted] (3.4,-2) -- (3.4,2);
        \draw[blue] (2.2,0.2) -- (3.3,0.2);
        \filldraw[blue] (2.7,0.2) circle (0pt) node[anchor=north]{$2R_{overlap}$};
        \filldraw[black] (3.4,-2) circle (0pt) node[anchor=north]{$\vec{y}_0$};
        \filldraw[red] (3.4,-1) circle (0pt) node[anchor=north]{$t'_1$};
        \filldraw[red] (3.4,1.4) circle (0pt) node[anchor=north]{$t'_2$};
        \end{tikzpicture}
        \caption{Same results hold as in Fig. \ref{fig:R-y<y0<y'} by symmetry between $y=0$ and $y=y_0$.}
        \label{fig:R-y'<y0<y}
    \end{minipage}\hfill
    \begin{minipage}{0.48\textwidth}
        \centering
        \begin{tikzpicture}
        \draw[black, thick] (0,0) -- (2,2) -- (4,0) -- (2,-2) -- (0,0);
        \draw[black, dotted] (2,-2) -- (2,2);
        \draw[line width=0.01,red] (0.4,-0.2) -- (2,1.4) -- (3.6,-0.2) -- (2,-1.8) -- (0.4,-0.2);
        \filldraw[black] (2,-2) circle (0pt) node[anchor=north]{$\Vec{0}$};
        \filldraw[red] (2,-1.8) circle (0pt) node[anchor=south]{$t_1$};
        \filldraw[red] (2,1.4) circle (0pt) node[anchor=north]{$t_2$};
        
        \draw[black, thick] (1.4,0) -- (3.4,2) -- (5.4,0) -- (3.4,-2) -- (1.4,0);
        \draw[line width=0.01,red] (1.7,0.1) -- (3.4,1.8) -- (5.1,0.1) -- (3.4,-1.6) -- (1.7,0.1);
        \draw[black, dotted] (3.4,-2) -- (3.4,2);
        \draw[blue] (1.7,0.1) -- (3.3,0.1);
        \filldraw[blue] (2.7,0.1) circle (0pt) node[anchor=north]{$2R_{overlap}$};
        \filldraw[black] (3.4,-2) circle (0pt) node[anchor=north]{$\vec{y}_0$};
        \filldraw[red] (3.4,-1.6) circle (0pt) node[anchor=south]{$t'_1$};
        \filldraw[red] (3.4,1.8) circle (0pt) node[anchor=north]{$t'_2$};
        \end{tikzpicture}
        \caption{The number of trajectories enclosed in the maximal overlap depends on conditions.}
        \label{fig:R-y0<yy'}
    \end{minipage}
\end{figure}

\noindent\textbf{\large{C.2 When $y_{int}>y_0$}}
\vspace{2mm}

See Fig. \ref{fig:R-y0<y} ($y_{int}<R$ holds because $t_1>-\infty$ and $t_2<\infty$), the conditions on $t_1', t_2'$ are determined by solving $y'_{int}=y_{int}-y_0$ with $\eta_1+\eta_2=\eta_1'+\eta_2'$. We find the conditions for overlap to exist is
\begin{equation}
    \begin{split}
         t_2' &> t_a' = -R\ln\bigg(e^{-t_1/R}-\frac{y_0}{R}\bigg) \hspace{1mm},\\
        \text{and} \hspace{5mm} t'_1 &< t_b' = - R\ln\bigg(e^{-t_2/R}+\frac{y_0}{R}\bigg) \hspace{1mm}.
    \end{split}
\end{equation}

\textbf{C.2.1 When $y'_{int}\leq y_0$} 
\vspace{2mm}

See Fig. \ref{fig:R-y'<y0<y}. By symmetry of the geometry, the results, in this case, are exactly the same as in case when $y'_{int}>y_0$, except that the deficit angle is $\delta'$ instead of $\delta$.

\vspace{2mm}
\textbf{C.2.2 When $y'_{int}>y_0$}
\vspace{2mm}

See Fig. \ref{fig:R-y0<yy'}. Whether there are one or two deficit angles depends on the situation. Here $y_{int}\big((\eta_1'+\eta_2')/2\big)$ is the radius of the $(t_1, t_2)$ causal diamond evaluated at time $(\eta_1'+\eta_2')/2$, similar to the definition of $y_{int}'\big((\eta_1+\eta_2)/2\big)$.

\textbf{a}. If $\eta_1+\eta_2=\eta_1'+\eta_2'$, the entropy of overlapping causal diamond is $(2\pi-\delta-\delta')R_{overlap}$. 

\textbf{b}. If $\eta_1+\eta_2\neq \eta_1'+\eta_2'$ and $max\{y_{int}'((\eta_1+\eta_2)/2), y_{int}((\eta_1'+\eta_2')/2)\}\leq y_0$, the entropy is $2\pi R_{overlap}$.

\textbf{c}. If $\eta_1+\eta_2\neq \eta_1'+\eta_2'$, $max\{y_{int}'((\eta_1+\eta_2)/2), y_{int}((\eta_1'+\eta_2')/2)\} > y_0$ and $min\{y_{int}'((\eta_1+\eta_2)/2), y_{int}((\eta_1'+\eta_2')/2)\}\leq y_0$, the entropy is $(2\pi-\{\delta, \delta'\}_{min})R_{overlap}$ where the deficit angle given by
\begin{equation}
    \{\delta, \delta'\}_{min} = \begin{cases}
        \delta \hspace{5mm} \text{if} \hspace{2mm} min\{y_{int}'((\eta_1+\eta_2)/2), y_{int}((\eta_1'+\eta_2')/2)\} = y_{int}((\eta_1'+\eta_2')/2) \hspace{1mm},\\
        \delta' \hspace{5mm} \text{if} \hspace{2mm} min\{y_{int}'((\eta_1+\eta_2)/2), y_{int}((\eta_1'+\eta_2')/2)\} = y_{int}'((\eta_1+\eta_2)/2) \hspace{1mm}.
    \end{cases}
\end{equation}

\textbf{d}. If $\eta_1+\eta_2\neq \eta_1'+\eta_2'$ and $min\{y_{int}'((\eta_1+\eta_2)/2), y_{int}((\eta_1'+\eta_2')/2)\} > y_0$, the entropy is $(2\pi-\delta-\delta')R_{overlap}$. 

\vspace{5mm}

To conclude, the radius of the overlapping causal diamond is always given by Eq.(\ref{eq:roverlap}) which is a function of $y_0, t_1, t_2, t_1', t_2'$. The deficit angle $\Delta$ takes value of $0, \delta, \delta', \delta+\delta'$, which value it takes depends on the relation among $y_0, t_1, t_2, t_1', t_2'$. The entropy is the area of the bifurcation surface of the overlapping diamond which is $(2\pi-\Delta)R_{overlap}$.

%We thank ... The work of TB is partially supported by the Department of Energy under grant DOE SC0010008.  The work of KZ is supported by the Heising-Simons Foundation ``Observational Signatures of Quantum Gravity'' collaboration grant 2021-2817, by the DoE under contract DE-SC0011632, and by a Simons Investigator award.

\def\bibsection{\section*{\refname}} 
\bibliographystyle{unsrtnat}
\bibliography{references}

\end{document}